\documentclass[manuscript]{aastex62}

\usepackage{xcolor}

\received{\today}
\submitjournal{ApJ}

\shorttitle{PVECs and HXR Sources in Solar Flares}
\shortauthors{Zimovets et al.}

\begin{document}

\title{Relationships between Photospheric Vertical Electric Currents and Hard X-Ray Sources in Solar Flares: Statistical Study}

\correspondingauthor{I.V. Zimovets}
\email{ivanzim@iki.rssi.ru}

\author[0000-0001-6995-3684]{I.V. Zimovets}
\affiliation{Space Research Institute of the Russian Academy of Sciences (IKI), 84/32 Profsoyuznaya Str, Moscow, Russia, 117997}

\author[0000-0002-5719-2352]{I.N. Sharykin}
\affiliation{Space Research Institute of the Russian Academy of Sciences (IKI), 84/32 Profsoyuznaya Str, Moscow, Russia, 117997}

\author{W.Q. Gan}
\affiliation{Key Laboratory of Dark Matter and Space Astronomy, Purple Mountain Observatory, Chinese Academy of Sciences, Nanjing, China, 210034}


\begin{abstract}

There are still debates whether particle acceleration in solar flares may occur due to interruption of electric currents flowing along magnetic loops. To contribute to this problem, we performed the first statistical study of relationships between flare hard X-ray (HXR; $50-100$ keV) sources observed by the \textit{Ramaty High-Energy Solar Spectroscopic Imager} (RHESSI) and photospheric vertical electric currents (PVECs, $j_{r}$) calculated using vector magnetograms obtained with the Helioseismic and Magnetic Imager (HMI) on-board the \textit{Solar Dynamics Observatory} (SDO). A sample of 48 flares, from C3.0 to X3.1 class, observed in central part of the solar disk by both instruments in 2010--2015 was analyzed. We found that $\approx 70$\% of all HXR sources overlapped with islands or ribbons of enhanced ($\left| j_{r} \right| \gtrsim 10^{4}$ statampere~cm$^{-2}$) PVECs. However, less than $\approx 40$\% of the HXR sources overlapped with PVEC maxima, with an accuracy of $\pm 3^{\prime\prime}$. More than in half of the flares there were HXR sources outside regions of enhanced PVECs. We found no correlation between intensity of the HXR sources and PVEC density or total PVEC under them. No systematic dissipation of PVECs under the HXR sources was found during the flares. Collectively, the results do not support the current-interruption flare models. However, the results indicate the importance of the presence of longitudinal currents in flare regions. Understanding of their specific role in the processes of energy release, plasma heating, and acceleration of particles requires further investigation.  

\end{abstract}

\section{INTRODUCTION} 
\label{S-Intro}

It is generally accepted that solar flares are the result of explosive release of free magnetic energy stored in active regions in the form of electric currents \citep[e.g.][]{Schmieder2018,Fleishman2018}. However, there are still active debates on how exactly the transformation of free magnetic energy into other energy channels, such as kinetic energy of charged particles, electromagnetic radiation and plasma waves, occurs in flare regions. The most common concept is that the flare energy release and particle acceleration takes place in coronal current sheets as a result of magnetic reconnection \citep{Priest2002,Somov2013}. There are a large number of observations supporting this concept \citep[e.g.][]{Benz2008,Krucker2008,Su2013}. 

Nevertheless, there are alternative concepts. One of them is the concept that energy can be explosively released as a result of interruption of currents flowing along magnetic loops \citep{Alfven1967,Spicer1981,Zaitsev2008}. Models based on this concept are called the current-interruption models. Despite the fact that these models cannot easily explain some of the observable properties of flares, in particular, the `above-the-loop-top' hard X-ray (HXR) sources \citep{Masuda1994,Krucker2008}, it has an important merit. Namely, it copes with the solution of the so-called `number problem' of accelerated particles, since in the framework of this model particle acceleration can occur in the chromosphere, in a region with a sufficiently high plasma density \citep{Zaitsev2015,Zaitsev2016a}.

There have been many attempts to test the current-interruption models. The main approach is as follows. The position of flare sources observed mainly in H$\alpha$ (or H$\beta$) or HXR emissions was compared with maps of photospheric vertical electric currents (PVECs, $j_{r}$), which were calculated on the basis of Ampere's law using photospheric vector magnetograms. If position of emission sources coincided with regions of enhanced PVECs exceeding a certain level (usually $1 - 3 \sigma \left( j_{r} \right)$ of the background, calculated from quiet areas of the Sun), then it was concluded that the model satisfies the observations. For the first time, this approach was systematically implemented by \citet{Moreton1968}. Based on an analysis of 30 flares that occurred in one sunspot group over 8 days, it was found that 80\% of flare H$\alpha$ knots coincided, within accuracy of $6^{\prime\prime}$, with strong ($\left| j_{r} \right| \geq 2.5 \times 10^{3}$ statampere cm$^{-2}$) PVECs. Similar result was obtained by \citet{Zvereva1970}, who, based on an analysis of two active regions that produced two `proton' flares, found that, within the same accuracy of $6^{\prime\prime}$, at least 74\% of all first flare H$\alpha$ brightenings coincide with the locations of the relative maximum of PVECs. Later, \citet{Lin1987} confirmed these results based on a single flare analysis. It was shown that H$\alpha$ kernels coincided with PVECs maximums within accuracy of $2^{\prime\prime}$. \citet{Romanov1990} analyzed 3 flares and found that some H$\alpha$ kernels were in PVECs maximums, some -- in periphery or between PVECs of opposite signs, some -- outside strong ($\left|j_{r} \right| \geq 10^{3}$ statampere cm$^{-2}$) PVECs. \citet{Abramenko1991} examined the observations of 2 active regions for 6 and 7 days and found that H$\alpha$ flare knots appeared most frequently in places with strong ($\left|j_{r} \right| \geq 3 \times 10^{2}$ statampere cm$^{-2}$) PVECs. In these works, a general conclusion was drawn that the observations correspond to the predictions of the current-interruption models. It is worth noting that positions of flare sources were obtained on the base of H$\alpha$ filtergrams. It is, however, known that flare brightnings observed by this way could not be necessary caused by precipitating energetic electrons, but also by thermal conduction from the overlying corona and high coronal pressure \citep{Canfield1984,Gan1991,Gan1992}. 

For this reason, in a series of works \citep{Canfield1993,deLaBeaujardiere1993,Leka1993}, spectrally resolved observations in H$\alpha$ were used and the positions of flare sources with specific line profiles, which could be caused by precipitating electrons, were determined for five flares in two active regions with accuracy of $\approx 3^{\prime\prime}$. The main finding of these works was that the sites of precipitation of energetic electrons to the chromosphere were on the shoulders of strong PVEC channels, rather than at PVEC maxima. It was argued that `these observations do not support a current-interruption model, unless the relevant currents are primarily horizontal'. An additional weighty argument in favor of this opinion was obtained by \citet{Li1997}. The advantage of this work was that instead of H$\alpha$ observations of the regions of precipitation of accelerated electrons, the data on HXR emission obtained with the \textit{Yohkoh} Hard X-ray Telescope \citep[HXT,][]{Kosugi1991a} were used. Observations in the HXR range provide more direct and reliable information about the region of interaction of accelerated electrons with dense chromospheric plasma. Based on the analysis of 6 solar flares in different active regions, observed with accuracy of $4^{\prime\prime}-6^{\prime\prime}$, it was confirmed that places of precipitation of energetic electrons preferentially occur adjacent to strong ($\left|j_{r} \right| \geq \left(0.9-3.2\right) \times 10^{3}$ statampere cm$^{-2}$) PVEC regions, but not in their maxima. It was also found that more intense conjugate footpoint HXR sources were emitted from regions of weaker magnetic field and PVEC. \citet{Li1997} concluded that their observations are not consistent with the current-interruption models but are in agreement with the `cornucopia' magnetic configuration of a flare region, where energetic electrons are reflected by a magnetic mirror in footpoints with stronger magnetic field.

After this, several more works were done to study relationships between flare emission sources and PVECs. \citet{Zhang1997} found the proximity of H$\beta$ sources to enhanced PVEC regions in one $\delta$-active region, although the emission sources were outside PVEC maxima. The observations were interpreted by the interaction of current-carrying loops. \citet{Ji2003} made a statistical analysis of relative spatial positions of H$\alpha$/H$\beta$ flare kernels and PVECs (and also photospheric horizontal electric currents, PHECs) in 79 solar flares observed in three active regions. For PVECs and PHECs the rates of `close correlation' were 29\% and 10\%, respectively, and the rates of `quasi-close correlation' were 50\% and 30\%. The `close correlation' and 'quasi-close correlation' means that a flare kernel is partially or completely overlapping with the 90\% and 80\% maximum isopleths, respectively, of an enhanced PVEC (or PHEC) region. It was also found that some flare kernels are correlated with both PVECs and PHECs, but most kernels are correlated with only one kind of photospheric currents, and only $\approx 6$\% of kernels are not correlated with either kind of currents. 

\citet{Sharykin2014a} found that some parts of a very fine ($\approx 0.1^{\prime\prime}$) H$\alpha$ ribbon observed in a C2.1 flare were superposed with PVEC maxima, while other parts of the ribbon were on the periphery of the strong PVEC region at the same time. In another weak C7.0 flare accompanied by a sunquake, \citet{Sharykin2015a} found good spatial coincidence (within $3^{\prime\prime}$) between a maximum of PVECs ($j_{r} \approx 7.8 \times 10^{4}$ statampere cm$^{-2}$) and a less intense flare HXR footpoint, as well as with a sunquake source, whereas an opposite more intense conjugate HXR footpoint was outside strong PVECs. 

An important contribution was made by \citet{Musset2015}, who studied relations between HXR sources and PVECs in the famous powerful X2.2 flare on 15 February 2011 using the \textit{Ramaty High-Energy Solar Spectroscopic Imager} \citep[RHESSI, ][]{Lin2002} observations and photospheric vector magnetograms constructed with the observational data by Helioseismic and Magnetic Imager \citep[HMI, ][]{Scherrer2012} on-board the \textit{Solar Dynamics Observatory} \citep[SDO, ][]{Pesnell2012}. They found that some of the HXR sources appeared on $j_{r}$-ribbons pre-existed before the flare. More interestingly, they discovered the appearance of new HXR (50--100 keV) sources in the same places with the appearance of new PVECs at the same time interval. Local increase in PVECs during the flare impulsive phase has been also reported in several other studies \citep[e.g., ][]{Janvier2014,Sharykin2014a,Sharykin2015,Sharykin2015a,Janvier2016,Sharykin2019}. Similar to \citet{Janvier2014}, \citet{Musset2015} interpreted their observational results in the framework of the scenario, according to which the acceleration of electrons and a local increase in PVECs is a consequence of magnetic reconnection in a coronal current sheet. \citet{Tan2006} studied two different flares and found different behavior of PVECs in them. PVEC density, $j_{r}$, dropped rapidly near the flaring neutral line around the onset of the compact flare, while $j_{r}$ increased continuously with continuously emerging magnetic flux just before and during the bigger two-ribbon flare. \citet{Tan2006} offered a possible explanation for the found difference: magnetic reconnection could happen at different heights in the two events, near the photosphere for the first flare, and higher up for the second flare. 
 
Let's briefly summarize the aforementioned results of studies on connections between flare emission sources and PVECs. (1) There is a general tendency for flare sources to appear near enhanced PVEC regions. (2) Different studies show different correspondence between the position of flare sources and PVECs. The highest percentage of intersections (up to 80\%) is for flare kernels observed using H$\alpha$ filtergrams. These observations, however, do not guarantee that the observed sources are the result of interaction of accelerated electrons with dense plasma in footpoints of flare loops. (3) Spectrally resolved H$\alpha$ observations of five flares taking this circumstance into account, as well as the observations of six flares in the HXR range, carried out in the 1990s, showed that the flare sources tend to be located on the periphery of the regions of strong PVECs and to avoid their maxima. (4) It must be borne in mind that PVEC maps were obtained on the basis of observations made far from simultaneously with the observations of flare sources. The time difference in some cases reached several hours. Taking into account recent observations of fast variations in PVECs in the impulsive phase of several flares, the results should be treated with caution. (5) Unlike observations of H$\alpha$/H$\beta$ flare sources, no statistical study of the relationships (both for spatial position and amplitude) of flare HXR sources and PVECs has been done. There were only several case studies considered single flares or a set of a few ($N \leq 6$) flares. This makes generalization difficult. 

To fulfill this gap, in this article, we present the first statistical study of the relationships between flare HXR sources and PVECs. This study is based on the observations of the Sun in the HXR range with high spatial (up to $2.26^{\prime\prime}$) and temporal (up to 4 s) resolution by RHESSI, in conjunction with the photospheric vector magnetograms obtained continuously with the HMI/SDO observational data with a time cadence of 12 min and high angular resolution of $\approx 1^{\prime\prime}$, in the 24$^{th}$ solar cycle. 

The article is organized as follows. In Section~\ref{S-Data} we present descriptions of the observational data used and methodology of its analysis. The data analysis is performed in Section~\ref{S-Results}. In Section~\ref{S-Discussion} the results of the data analysis are summarized and discussed. The conclusion of the work is given in Section~\ref{S-Conclusions}.  

\section{DATA AND METHODOLOGY} 
\label{S-Data}

\subsection{HARD X-RAY DATA}
\label{Ss-HXRdata}

To study HXR sources of a set of solar flares we used data obtained with RHESSI \citep{Lin2002}. This space instrument detected photons in a broad energy range from a few keV to several MeV. RHESSI operated from February 2002 till April 2018. During that time it detected more than 120000 flares, tabulated information on which is contained in the RHESSI Flare Catalog. We used this catalog to select solar flares with the Helioprojective Cartesian (HPC) coordinates $-600^{\prime\prime} \leq \left[x_{f}, y_{f}\right] \leq +600^{\prime\prime}$, i.e. the flares selected were located not far from the solar disk center. Their Stonyhurst Heliographic (HG) longitudes and latitudes were within the range of $-40^{\circ} < [\phi_{f}, \theta_{f}] < +40^{\circ}$. This restriction allowed to minimize problems of determining pre-flare and post-flare maps of photospheric magnetic fields and PVECs, as well as of combining them with maps of chromospheric HXR sources (see Section~\ref{Ss-combdata}). Additional flare selection criterion was the detection by RHESSI of significant fluxes of HXR emission with energies $\geq 50$ keV. It is known that the HXR emission with such energies are mainly emitted from the chromospheric flare loop footpoints, with a small contribution from coronal HXR sources \citep[e.g., ][]{Fletcher2011}. We decided not to study HXR sources with energies below 50 keV, since there is higher probability that these HXR sources were mainly located in coronal parts of flare loops, at least in some events \citep{Veronig2004,Veronig2005a}. 

Applying the two criteria stated above to the RHESSI Flare Catalog in the time interval from May 2010 till December 2017, 132 events were initially found. The beginning of the interval is determined by the beginning of the receipt of regular observational data from HMI/SDO. We checked these 132 events and excluded those of them in which only a small part of the flare impulsive phase was observed, there was very strong noise, or the HXR flux in the range of 50--100 keV was too small to construct at least one high-quality image of the flare region. After such sifting, we have the final set of 48 solar flares for the analysis. The information on them is presented in Table~\ref{tab:flareinfo}. Among the selected flares, 6 flares were of X class, 36 -- of M class, and 6 -- of C class. The flares occurred in 31 different active regions.

The impulsive phase of many solar flares is a sequence of HXR peaks of duration from a fraction of second to several tens of seconds \citep{Dennis1988,Aschwanden2002}. Moreover, it is known that the sources of individual HXR peaks can be located in different places, usually in footpoints of different flux tubes organized in magnetic arcades and/or more complex structures, like magnetic flux ropes \citep[e.g.,][]{Fletcher2002,Krucker2003,Kuznetsov2016,Zimovets2018}. Since RHESSI rotated with a period of 4 s, and usually one needs to integrate over several RHESSI rotational periods to accumulate more detected HXR photons, we could synthesize HXR images and identify positions of HXR sources only for the strongest HXR peaks lasting longer than at least 8 s. For all 48 selected flares we identified time intervals of the strongest HXR peaks, avoiding times when the state of the RHESSI's attenuators was changed. In total, we selected 81 time intervals lasting from 8 to 92 s. The durations of the intervals were determined by the photon flux in the energy range of 50--100 keV. The number of HXR photons detected by one RHESSI's detector in each selected interval should be more than several hundreds. 

For each selected time interval, we synthesized 50--100 keV HXR maps of a flare region using two different algorithms, CLEAN and PIXON \citep{Metcalf1996,Hurford2002}, widely used in studies of solar flares. We decided to use two different algorithms to test the effect of the methodology on the final results. When using the CLEAN algorithm, we mainly used data from the RHESSI's detectors (sub-collimators) 2--8, sometimes adding data from detector 1 with the finest sub-collimator for a compact flare region and a strong HXR flux. The ``natural weighting'' was applied for the different sub-collimators. Data from all 9 detectors were mostly used to construct images using the PIXON algorithm, since this algorithm decides automatically which data to select. The data of only the frontal segments of the RHESSI's detectors was used. The pixel size of the synthesized images was $1^{\prime\prime}$ or $2^{\prime\prime}$, depending on a size of a flare region analyzed. Virtually all 50--100 keV HXR sources of a given flare must be contained in a synthesized map.  

\begin{figure}[!t]
\centering
\centerline{\includegraphics[width=0.9\linewidth]{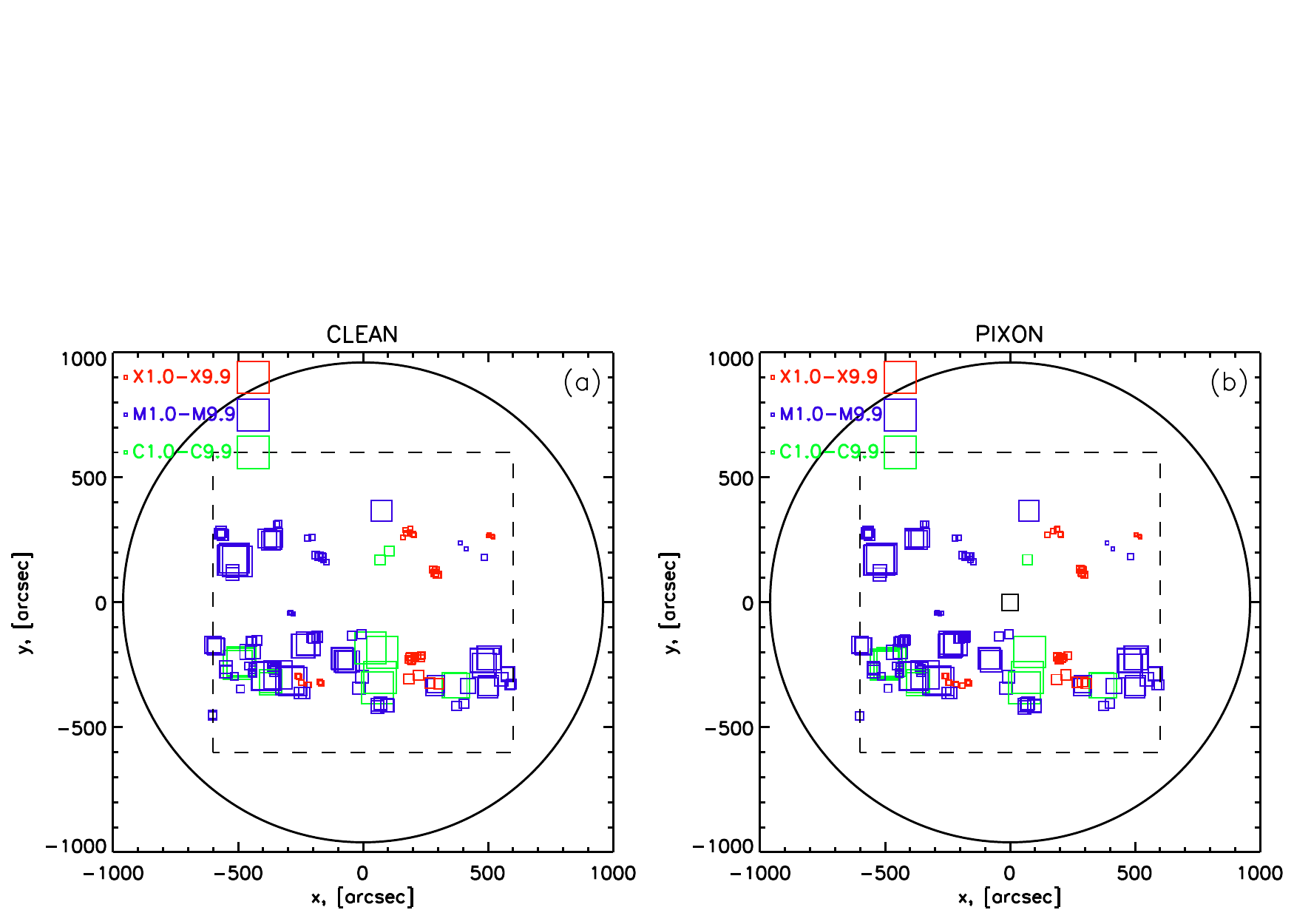}}
\caption{Positions of the brightest pixels of all HXR (50--100 keV) sources reconstructed with the CLEAN (a) and PIXON (b) algorithms for 48 solar flares studied. The X-ray class of the corresponding solar flares is denoted by squares of the appropriate size and color (green, blue, red --- C, M, X classes, respectively). The optical solar limb is shown by the bold circle. The region for initial selection of the flares ($-600^{\prime\prime} \leq \left[x_{f}, y_{f}\right] \leq +600^{\prime\prime}$) is shown by the dashed square. }
\label{all_hxrc_posit_fig}
\end{figure}

For all 48 flares and 81 time intervals selected we identified 177 and 186 HXR sources on the images reconstructed with the CLEAN and PIXON algorithms, respectively. Each HXR source is defined as an isolated cluster of bright pixels around the locally brightest one. The number of HXR sources found in the images of CLEAN and PIXON for each considered time interval is shown in columns 7 and 8, respectively, in Table~\ref{tab:flareinfo}. There was only one HXR source in 8 ($\approx 10$\%) and 13 ($\approx 16$\%) time intervals for CLEAN and PIXON, respectively; simultaneously two HXR sources were in 57 ($\approx 70$\%) and 46 ($\approx 57$\%) time intervals for CLEAN and PIXON, respectively; three HXR sources --- in 9 ($\approx 11$\%) time intervals both for CLEAN and PIXON; four HXR sources --- in 7 ($\approx 9$\%) and 11 ($\approx 14$\%), respectively; and five HXR sources --- in 2 ($\approx 2$\%) time intervals for PIXON only. Positions of the brightest pixels of all HXR sources reconstructed with the CLEAN and PIXON algorithms are shown in Figure~\ref{all_hxrc_posit_fig} together with the notations of an X-ray class of a corresponding solar flare. It is interesting to note that most of the HXR sources were observed in the southern hemisphere, where 33 ($\approx 69$\%) investigated flares occurred. The prevalence of solar flares in the southern hemisphere in 2002--2017 has been reported by \citet{Abdel-Sattar2018}.

\subsection{MAGNETIC FIELD DATA}
\label{Ss-magdata}
 
To construct maps of PVECs we used the photospheric vector magnetograms produced with the observational data of HMI/SDO. More specifically, we used the Spaceweather HMI Active Region Patches (\texttt{SHARP}) data series \citep{Bobra2014,Hoeksema2014}. \texttt{SHARP} data contains several different space-weather quantities calculated from the photospheric vector magnetograms and 31 data segments, including three components of the vector magnetic field, optical continuum intensity, Doppler velocity, error maps, etc. In particular, in the \texttt{hmi.sharp\_cea\_720s} fits-files, which we used, the magnetic field vector, $\textbf{B}$, is remapped to a Lambert Cylindrical Equal-Area (CEA) projection \citep{Thompson2006} and decomposed into three magnetic components in the spherical coordinate system: $\left(B_{r}, B_{\phi}, B_{\theta}\right)$. It is important to note that the azimuthal component of the vector magnetic field was disambiguated using the Minimum Energy Code (ME0) to resolve the $180^{\circ}$ ambiguity. A confidence level of disambiguation is contained in the \texttt{conf\_disambig} segment. We avoided pixels with the non-disambiguated magnetic field. The \texttt{SHARP} CEA pixels have a linear dimension in the x-direction of 0.03 heliographic degrees in the rotated coordinate system and a fixed area on the photosphere of $1.33 \times 10^{5}$ km$^{2}$. The data segments contained in \texttt{SHARP}s are partial-disk, automatically-identified active region patches. They are calculated every 12 minutes. 

For each selected event we used two sets of \texttt{SHARP} data --- one for a time just before a flare impulsive phase and one for a time just after it. We identified a flare impulsive phase as a time interval with count rates in the RHESSI 25--50 keV channel exceeding the background level. Using of ``pre-flare'' and ``post-flare'' magnetograms helps to avoid disturbances of magnetic field measurements due to enhanced emission caused by precipitating energetic particles and powerful heat fluxes arising in the flare impulsive phase \citep[e.g., ][]{Sun2017}. Further, in the text and in figures, the pre-flare and post-flare characteristics will be denoted by the subscripts $0$ and $1$, respectively.

\subsection{CALCULATION OF PVEC AND COMBINATION WITH HXR MAPS}
\label{Ss-combdata}

For each event, firstly, we converted the CEA coordinates of pre-flare and post-flare \texttt{SHARP} vector magnetograms to the HG coordinates. We also converted the HPC coordinates of the flare HXR maps, obtained with the RHESSI data, to the HG coordinates. Secondly, we differentially rotated the HG coordinates of the post-flare magnetograms and flare HXR maps to the time of the pre-flare magnetograms. Then, we converted the HG coordinates to the spherical coordinates. We calculated the photospheric vertical (i.e., radial) electric current (PVEC) density in the spherical coordinates using the circulation theorem of magnetic field induction (Ampere's law) in the differential form: 
\begin{equation}
 j_{r} \left(r=R_{s},\phi,\theta\right) = \frac{c}{4 \pi \mu} \left(\nabla \times \textbf{B} \right)_{r} \approx \frac{c}{4 \pi} \frac{1}{R_{s} \sin \theta} \left(\frac{\Delta B_{\varphi}}{\Delta \theta} \sin \theta + B_{\varphi} \cos \theta-\frac{\Delta B_{\theta}}{\Delta \varphi} \right),
\label{eq:jr}
\end{equation}
where $c$ is the speed of light in vacuum, $R_{s}$ is radius of the Sun, and the magnetic permeability  $\mu = 1$. After that, we calculated area of each pixel $\Delta S_{p}\left(r=R_{s},\phi,\theta\right)$ and photospheric vertical current $J_{r} \left(R_{s},\phi,\theta\right) = j_{r}\left(R_{s},\phi,\theta\right) \times \Delta S_{p}\left(R_{s},\phi,\theta\right)$ through a pixel. 

\begin{figure}[!t]
\centering
\centerline{\includegraphics[width=0.9\linewidth]{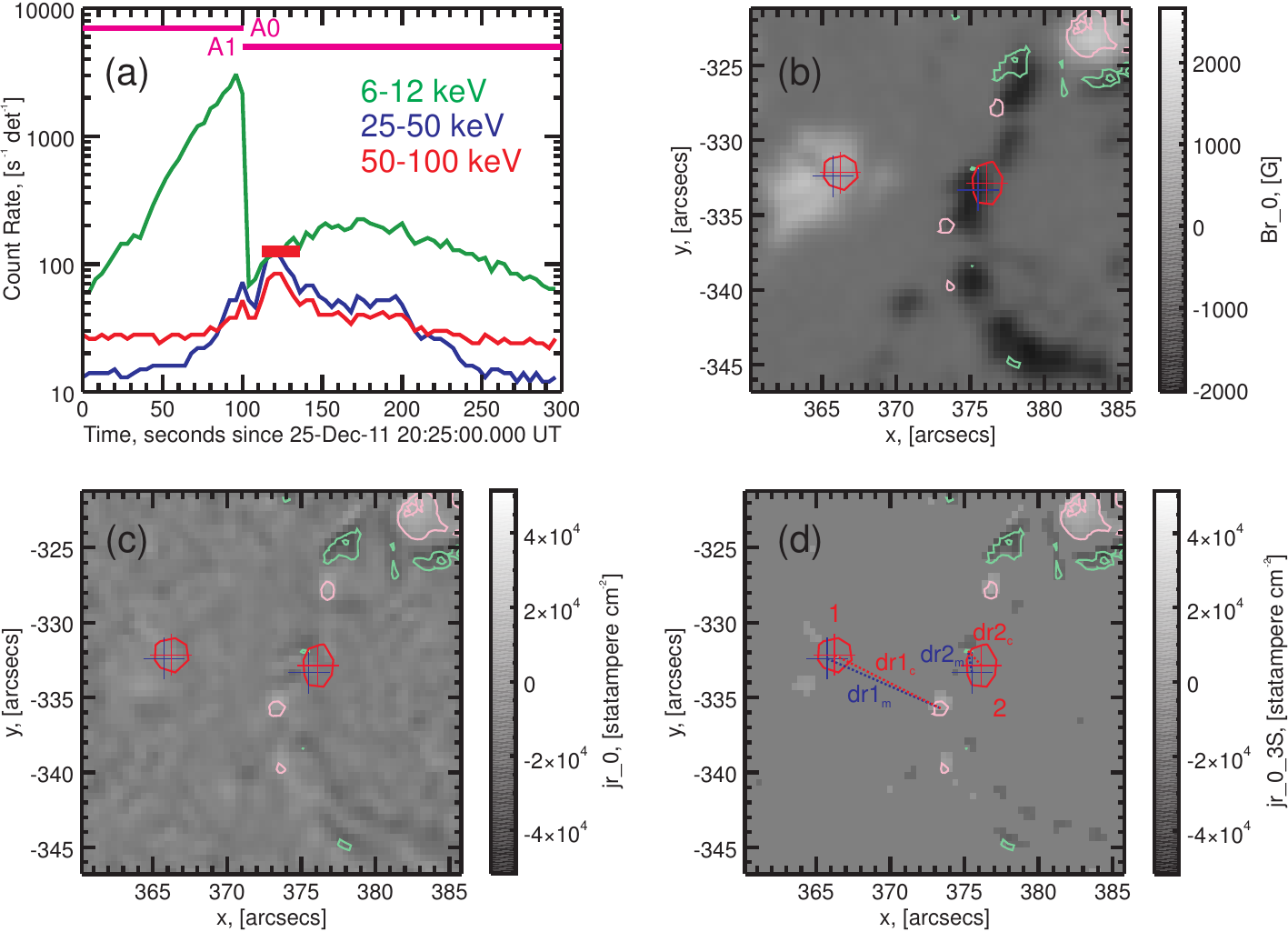}}
\caption{Example of one solar flare studied: the SOL2011-12-25T20:23 event. The panel (a) shows the standard 4-second RHESSI count rates in three energy channels 6--12 (green), 25--50 (blue), and 50--100 (red) keV. The pink horizontal lines above show the state of the RHESSI attenuators (A0, A1). The red thick horizontal bar above the 50--100 keV count rate indicates the time interval for which the HXR sources (shown on b--d) was synthesized. The background grayscale images on (b), (c), and (d) are the pre-flare maps of the radial magnetic field component $B_{r}$, ``original'' PVEC density $j_{r}$, and ``cleaned'' PVEC density above three standard deviations of the background, respectively. The colorbars are shown to the right of the figures. Positions of enhanced PVEC $j_{r}$-regions at levels of $1, 2, \ldots, 8 \times j_{r}^{thr}$ are shown by pink (positive) and cyan (negative) contours on (b--d). The 50--100 keV HXR sources at a level of 90\% of their maximum brightness, reconstructed with the CLEAN algorithm for the time interval marked with the red thick horizontal bar on (a), are shown by the red contours on (b--d). The blue and red crosses show positions of the centers of maximum brightness and ``centers-of-mass'' of brightness, respectively, of the HXR sources. The vertical and horizontal sizes of the crosses indicate the estimated errors in determining the HXR source positions, $\pm \sigma_{\text{HXR}}$. The distances between the positions of the maximum brightness ($dr1_{m}$ and $dr2_{m}$, blue) or ``center-of-mass'' of brightness ($dr1_{c}$ and $dr2_{m}$, red) of the HXR sources 1 or 2 and the closest local maxima of the nearest enhanced $j_{r}$-regions are shown by the dotted lines of corresponding colors on (d).}
\label{251211_flare_fig}
\end{figure}

The calculated PVEC maps contain significant noise. To determine the noise level, for each \texttt{SHARP} data set we selected a `background box' in a quiet Sun region without significant magnetic fields, constructed and plotted the distribution of $\left| j_{r} \right|$ for this `background box'. As shown in \citet{Zimovets2019}, for all the considered 48 active regions the $j_{r}$-distribution below the threshold value of $j^{thr}_{r} \approx 10^{4}$ statampere~cm$^{-2}$ can be well approximated by a Gaussian function, while it has a power-law shape above this threshold. It was argued that the Gaussian part of $j_{r}$-distribution represent data noise, but the power-law part can contain physically meaningful information. Using the least squares method, we fit the constructed $j_{r}$-distributions and obtained the standard deviation values, $\sigma\left(j_{r}\right)$, for each $j_{r}$-map. For all active regions studied, $2100 < \sigma\left(j_{r}\right) < 3200$ statampere~cm$^{-2}$. After that, we built ``cleaned'' $j_{r}$-maps (and $J_{r}$-maps), the values in pixels of which are equal to values of `original' $j_{r}$-maps if they exceed $3 \sigma\left(j_{r}\right)$, or equal to zero otherwise. In such ``cleaned'' maps, noise does not contribute to the estimate of $\left\langle j_{r}\right\rangle$ (here and below $\left\langle \ldots \right\rangle$ means averaging) or $J_{r}$ over the region under consideration, however, it gives underestimated values, since some pixels can have artificial zero values. We will use both ``original'' and ``cleaned'' $j_{r}$-maps for comparison. As an example, two such $j_{r}$-maps, for the SOL2011-12-25T20:23 event, are shown in Figure~\ref{251211_flare_fig}(c,d).

In order to combine HXR maps with the photospheric magnetograms and PVECs, and also for ease of visualization, we converted the spherical and HG coordinates to the HPC coordinates and interpolated all maps to the same uniform grids of HPC coordinates. We determined each HXR source as a cluster of pixels with an intensity of at least 90\% of the intensity of the brightest pixel in this cluster. Such a high level is chosen to limit the size of the HXR sources. HXR sources determined at lower levels usually have significantly larger spatial scale than PVECs. With a decrease in the level, an area of the HXR sources increases significantly and, when calculating PVECs, averaging or summing proceeds over a larger area. We decided to confine the study by analysis of the most central parts of the sources. For each HXR source we also determined positions of its center of maximum brightness and `center-of-mass' of brightness, together with a possible error. We estimated the error by the following way: 
\begin{equation}
\sigma_{\text{HXR}}=\sqrt{\text{FWHM}_{\text{HSI}}^{2}+\Delta p_{\text{HSI}}^{2}+\text{FWHM}_{\text{HMI}}^{2}+\Delta p_{\text{HMI}}^{2}+\Delta h^{2}}/2, 
\label{eq:hxrposerr}
\end{equation}
where $\text{FWHM}_{\text{HSI}}=2.26^{\prime\prime}$ or $=3.92^{\prime\prime}$ is the angular resolution (i.e. a full width at half maximum, FWHM) of the finest RHESSI's collimator used to synthesize a HXR map, $\Delta p_{\text{HSI}} = 1^{\prime\prime}$ or $=2^{\prime\prime}$ is the angular size of a HXR map pixel chosen, $\text{FWHM}_{\text{HMI}} = 1^{\prime\prime}$ is the angular resolution of the HMI/SDO instrument, $\Delta p_{\text{HMI}} = 0.5^{\prime\prime}$ is the average angular size of an HMI map pixel, and $\Delta h$ is the projection distance of a chromospheric HXR source center. We assume that all HXR sources were in the chromosphere at an altitude of $h=2.5$ Mm above the photosphere. This gives an upper estimate for $\Delta h \approx \left(h/R_{s}\right) \sqrt{x^{2}_{\text{HXR}}+y^{2}_{\text{HXR}}}$. An example of the location of HXR sources (with its center of maximum brightness and `center-of-mass' of brightness) on the pre-flare map of the vertical magnetic component, $B_{r0}$, as well as on the ``original'' and ``cleaned'' pre-flare $j_{r0}$-maps for the SOL2011-12-25T20:23 event is shown in Figure~\ref{251211_flare_fig}(b--d).

For each HXR source, the coordinates of all pixels satisfying the indicated criterion are determined. Based on this, various characteristics of the photospheric magnetic fields and PVECs in the area under each HXR source were calculated. In particular, we calculated the average, maximum, and minimum values of the radial, $B_{r}$, and tangential, $B_{k}$, magnetic components, and PVEC density, $j_{r}$, as well as the total current under a HXR source, $J_{r}$. The values of physical characteristics are calculated both taking into account their sign and without taking it into account, e.g. we calculated $\left|\left\langle j_{r} \right\rangle\right|$ and $\left\langle \left| j_{r} \right|\right\rangle$ separately. We also calculated the ratios of the minimum, maximum, average (and also total $J_{r}$) values of these physical characteristics within a HXR source area after and before the flare impulsive phase (e.g. $\left\langle \left| j_{r1} \right|\right\rangle/\left\langle \left| j_{r0} \right|\right\rangle$ or $\left| J_{r1} \right|/\left| J_{r0} \right|$) in order to check the presence of their systematic changes during the flares.

\section{DATA ANALYSIS AND RESULTS}
\label{S-Results}

\subsection{RELATIVE LOCATIONS OF HXR SOURCES AND ENHANCED PVEC REGIONS}
\label{Ss-RelLoc}

Firstly, we analyzed spatial location of the flare HXR sources relative to regions of enhanced pre-flare PVECs. By enhanced PVEC regions we called clusters of pixels with the same sign of PVEC satisfying the following criteria: 
\begin{equation}
\left| j_{r}\right | \geq j^{thr}_{r} \approx 10^{4} \text{   statampere~cm$^{-2}$.}
\label{eq:jthr}
\end{equation}
The selection of this threshold value is indicated above (see Section~\ref{Ss-combdata}). For all the active regions studied this value exceeds triple standard deviation of the background noise: $j^{thr}_{r} > 3 \sigma\left(j_{r}\right)$.  

We found that in 43 out of 48 (90\%) flares studied at least one HXR source, reconstructed both with CLEAN and PIXON, was in enhanced PVEC regions. This means that at least one of the following three criteria is met for at least one HXR source of a given flare: 1) an iso-contour at a level of 90\% of maximum HXR source brightness intersects with or completely lies inside an enhanced PVEC region; or an enhanced PVEC region overlaps, at least partially, with 2) a center of maximum HXR source brightness or 3) a center-of-mass of HXR source brightness, within the error determined by expression (\ref{eq:hxrposerr}). In 36 (75\%) and 29 (60\%) flares, at least one HXR source, constructed with CLEAN and PIXON, respectively, was in a local maximum of an enhanced $j_{r}$-region. In 31 (65\%) and 25 (52\%) flares at least one HXR source, constructed with CLEAN and PIXON, respectively, was in a global maximum of an enhanced $j_{r}$-region. In 11 (23\%) and 8 (17\%) flares at least one HXR source reconstructed with CLEAN and PIXON, respectively, was in major $j_{r}$-maxima of an entire parent active region. We need to clarify here that, according to our definition, an enhanced $j_{r}$-region can have one or several local maxima and only one global maximum. However, there were cases when a HXR source overlapped simultaneously with a few separate enhanced $j_{r}$-regions. In this case, the HXR source could overlap simultaneously with a few global $j_{r}$-maxima. Each active region has two major $j_{r}$-maxima, which correspond to the strongest positive (upward from the photosphere) and negative (down from the photosphere) $j_{r}$ peaks of an entire \texttt{SHARP} region.

In 17 (35\%) flares all HXRS constructed both with CLEAN and PIXON were in enhanced $j_{r}$-regions. In 7 (15\%) and 4 (8\%) flares all HXRS constructed with CLEAN and PIXON, respectively, overlapped with local maxima of enhanced $j_{r}$-regions. In 4 (8\%) and 3 (6\%) flares all HXRS constructed with CLEAN and PIXON, respectively, overlapped with global maxima of enhanced $j_{r}$-regions. Just in 1 (2\%) flare all HXRS constructed with CLEAN only overlapped with major $j_{r}$-maxima of an entire active region.   

We found that 130 out of 177 (73\%) and 125 out of 186 (67\%) HXR sources constructed with the CLEAN and PIXON algorithms, respectively, overlapped, at least partially, with enhanced $j_{r}$-regions. 75 (42\%) and 55 (30\%) HXR sources constructed with CLEAN and PIXON, respectively, overlapped with local maxima of enhanced $j_{r}$-regions. 54 (31\%) and 40 (22\%) HXR sources constructed with CLEAN and PIXON, respectively, overlapped with global maxima of enhanced $j_{r}$-regions. Only 16 (9\%) and 10 (5\%) HXR sources constructed with CLEAN and PIXON, respectively, overlapped with major maxima of enhanced $j_{r}$-regions of an entire parent active region. 

\begin{figure}[!t]
\centering
\centerline{\includegraphics[width=0.82\linewidth]{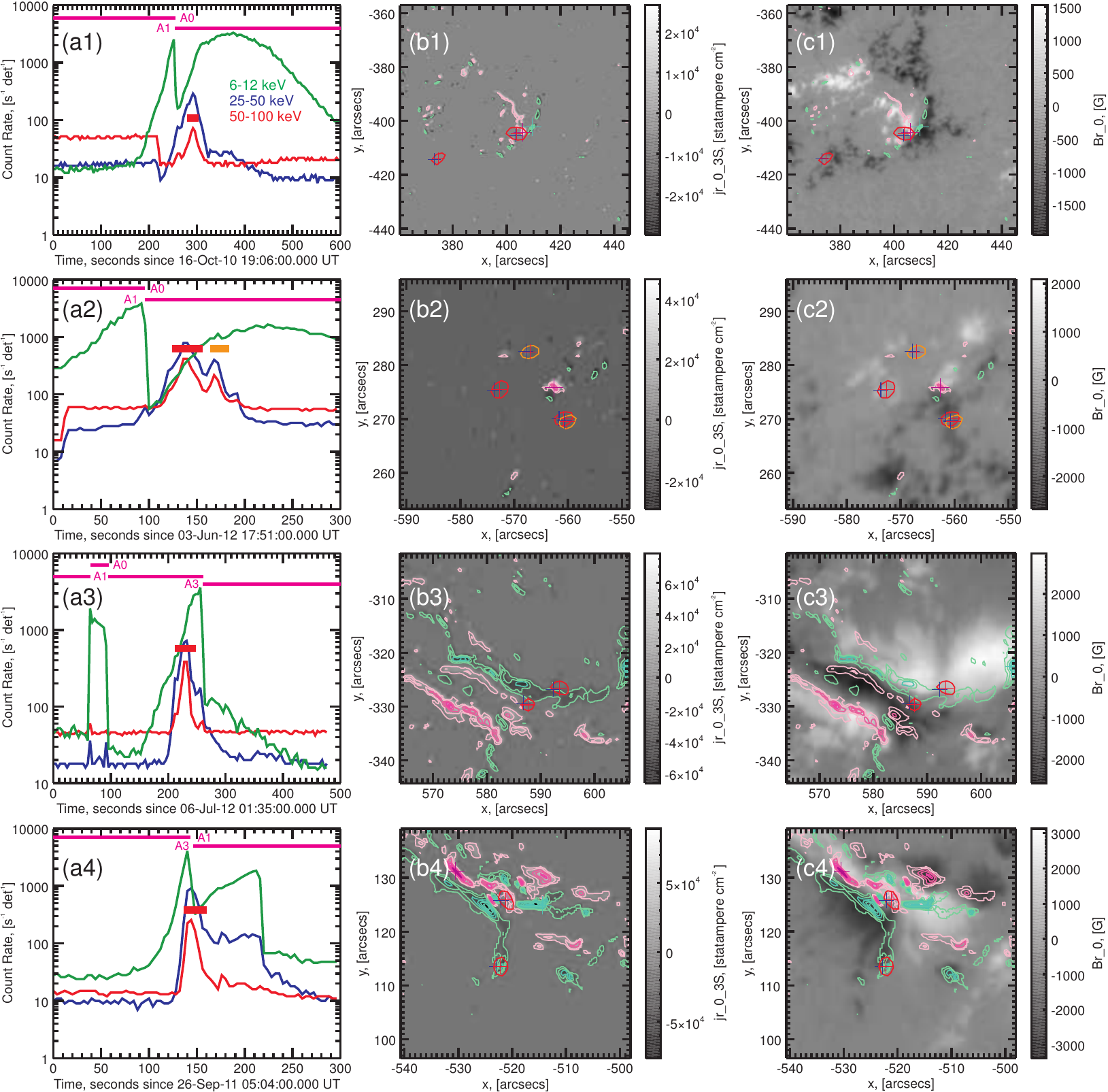}}
\caption{Examples of four investigated solar flares: SOL2010-10-16T19:07 (row 1), SOL2012-06-03T17:48 (row 2), SOL2012-07-06T01:37 (row 3), and SOL2011-09-26T22:12 (row 4). The left column (a) shows 4-second RHESSI count rates in three energy channels 6--12 (green), 25--50 (blue), and 50--100 (red) keV. The pink horizontal lines above show the state of the RHESSI attenuators (A0, A1, A3). The thick horizontal segments of different colors (red, orange) above the 50--100 keV count rates indicate time intervals for which images of HXR sources (shown on (b) and (c)) were constructed. The middle column (b) shows pre-flare maps of PVEC density $j_{r0}$ above three standard deviations of the background with the contour levels of $\pm 1, 2, \ldots, 8 \times j^{thr}_{r}$ (pink -- positive, cyan -- negative). Positions of the entire active region positive and negative $j_{r0}$ maxima are shown by the pink and cyan crosses, respectively. The 50--100 keV HXR sources, reconstructed with the CLEAN algorithm, at a level of 90\% of their maximum intensity, are shown by the contours of different colors corresponding to the time intervals of their appearance (shown on (a)). The blue and red crosses show positions of centers of maximum brightness and `centers-of-mass' of brightness of the HXR sources, respectively. The sizes of the crosses indicate the estimated errors, $\pm \sigma_{\text{HXR}}$, in determining the HXR source positions. The right panel (c) is similar to the middle panel (b), except that the background images on it represent pre-flare maps of the radial magnetic field component $B_{r0}$.}
\label{4pans_fig}
\end{figure}

There were various types of locations of HXR sources relative to enhanced $j_{r}$-regions. We divided all 48 events studied into four types. Four representative examples are shown in Figure~\ref{4pans_fig}. The most numerous are type I events (23 events, $\approx 48$\%), when one or several HXR sources overlapped with enhanced $j_{r}$-regions, while others were outside them at the same time (Figure~\ref{251211_flare_fig}(a1--c1)). Another example of such event is also shown in Figure~\ref{251211_flare_fig}. In 2 events of type I, all HXR sources were outside enhanced $j_{r}$-regions at some time intervals. We marked them as type Ia events. In 5 flares ($\approx 10$\%) of type II all HXR sources, reconstructed both with CLEAN and PIXON, were outside enhanced $j_{r}$-regions. One such example is shown in Figure~\ref{4pans_fig}(a2--c2). In 12 flares ($25$\%) of type III all HXR sources overlapped with enhanced $j_{r}$-regions, but not all HXR were in $j_{r}$-maxima (Figure~\ref{4pans_fig}(a3--c3)). In 8 other flares ($\approx 17$\%) all HXR sources of a flare overlapped with local or global maxima of enhanced $j_{r}$-regions (Figure~\ref{4pans_fig}(a4--c4)). These are type IV events. For several events, the analysis of images constructed with CLEAN and PIXON yielded different results. For definiteness, we used the results obtained with CLEAN here. The types of all events are indicated in the last column of Table~\ref{tab:flareinfo}.   

\subsection{RELATIVE LOCATIONS OF HXR SOURCES AND PVEC RIBBONS/ISLANDS}
\label{Ss-RelLocRibIsl}

All regions of enhanced PVECs can be conditionally divided into two groups: $j_{r}$-islands and $j_{r}$-ribbons. By $j_{r}$-islands we call more or less symmetric clusters of pixels with $\left| j_{r} \right| \geq j^{thr}_{r}$, and $j_{r}$-ribbons are structures of pixels with $\left| j_{r} \right| \geq j^{thr}_{r}$ elongated along a certain curve (usually not straight and approximately corresponding to the nearby photospheric magnetic polarity inversion line), whose length exceeds width at least three times. We found $j_{r}$-islands in parent active regions of all the flares studied, and $j_{r}$-ribbons in 43 ($\approx 90$\%) of active regions within $60^{\prime\prime}$ from the flare HXR sources. $j_{r}$-islands are much more common and numerous than $j_{r}$-ribbons. Usually, there are many small $j_{r}$-islands with an angular size of a few arc-seconds, and only a few $j_{r}$-ribbons longer than 10 arc-seconds. Representative examples of the flare regions with $j_{r}$-islands can be seen in Figures~\ref{251211_flare_fig}(c, d) and \ref{4pans_fig}(b2), and with $j_{r}$-ribbons in Figure~\ref{4pans_fig}(b1, b3, b4). Figures~\ref{4pans_fig}(b1, b3, b4) also show the presence of multiple $j_{r}$-islands in the flare regions containing $j_{r}$-ribbons.

\textbf{Overlapping of HXR sources with $j_{r}$-ribbons.} In 43 ($\approx 90$\%) active regions there were $j_{r}$-ribbons with and without HXR sources both for CLEAN and PIXON. 98 of all 177 ($\approx 55$\%) and 78 of all 186 ($\approx 42$\%) HXR sources, reconstructed with CLEAN and PIXON, respectively, overlapped with $j_{r}$-ribbons. 12 ($\approx 12$\%) and 5 ($\approx 6$\%), respectively, of these HXR sources overlapped with $j_{r}$-ribbons of mixed signs. 49 ($\approx 28$\%) and 27 ($\approx 15$\%), respectively, of all HXR sources overlapped with local maxima of $j_{r}$-ribbons. 21 ($\approx 12$\%) and 11 ($\approx 6$\%), respectively, of all HXR sources overlapped with global maxima of $j_{r}$-ribbons. 15 ($\approx 8$\%) and 10 ($\approx 5$\%), respectively, of all HXR sources overlapped with major maxima of $j_{r}$-ribbons, corresponded to major $j_{r}$-maxima of parent active regions.    

At least one HXR source overlapped with $j_{r}$-ribbons in 35 ($\approx 73$\%) and 32 ($\approx 67$\%) flares, for CLEAN and PIXON, respectively. At least one HXR source was in local maximum of $j_{r}$-ribbons in 25 ($\approx 52$\%) and 14 ($\approx 29$\%) flares for CLEAN and PIXON, respectively. At least one HXR source was in global maxima of $j_{r}$-ribbons in 16 ($\approx 33$\%) and 8 ($\approx 17$\%) flares for CLEAN and PIXON, respectively. At least one HXR source was in global maxima of $j_{r}$-ribbons, which were also the major $j_{r}$-maxima of an entire parent active region, in 10 ($\approx 21$\%) and 7 ($\approx 15$\%) flares for CLEAN and PIXON, respectively. 

\textbf{Overlapping of HXR sources with $j_{r}$-islands.} In all 48 (100\%) active regions there were $j_{r}$-islands with and without HXR sources reconstructed both with CLEAN and PIXON. 60 of all 177 ($\approx 34$\%) and 58 of all 186 ($\approx 31$\%) HXR sources, constructed with CLEAN and PIXON, respectively, overlapped with $j_{r}$-islands. 28 ($\approx 47$\%) and 23 ($\approx 40$\%), respectively, of these HXR sources overlapped with $j_{r}$-islands of mixed signs. 37 ($\approx 62$\%) and 33 ($\approx 57$\%), respectively, of these HXR sources overlapped with tiny $j_{r}$-islands. By a tiny island we mean a cluster of only 1--4 pixels of the same sign. We specifically noted such tiny $j_{r}$-islands, since their origin is under question. It is possible that they may represent not completely `cleaned' data noise. 33 ($\approx 19$\%) and 26 ($\approx 14$\%), respectively, of all HXR sources overlapped with local maxima of $j_{r}$-islands. The same is for overlapping with global maxima of $j_{r}$-ribbons. Only 1 of all HXR sources, both for CLEAN and PIXON, overlapped with the major maximum of $j_{r}$-island, corresponded to the major $j_{r}$-maximum of the parent active regions.

At least one HXR source overlapped with $j_{r}$-islands in 34 ($\approx 71$\%) and 35 ($\approx 73$\%) flares, for CLEAN and PIXON, respectively. At least one HXR source was in $j_{r}$-island local (and also global) maxima in 21 ($\approx 44$\%) and 19 ($\approx 40$\%) flares for CLEAN and PIXON, respectively.

In general, it can be concluded that the HXR sources overlapped with $j_{r}$-ribbons about 1.5 times more often than with $j_{r}$-islands. 

\subsection{DISTANCES BETWEEN HXR SOURCES AND ENHANCED PVEC REGIONS}
\label{Ss-RelLocRibIsl}

\begin{figure}[!t]
\centering
\centerline{\includegraphics[width=0.9\linewidth]{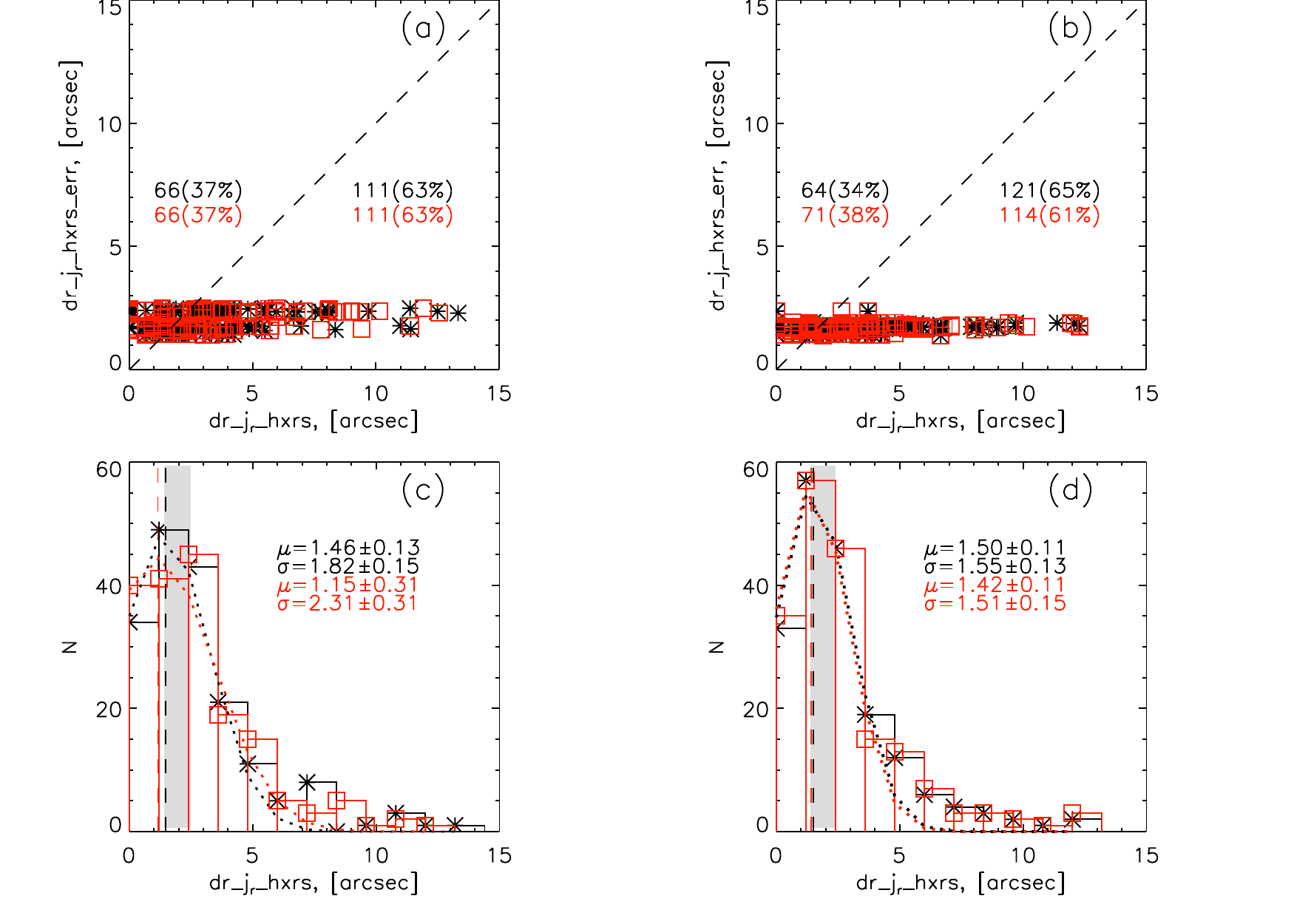}}
\caption{Results of measuring the distances between the flare HXR (50--100 keV) source centers and local maxima of pre-flare regions of enhanced PVEC density above the defined threshold level $\left|j_{r}\right| \geq j_{r}^{thr}$. The panels on the left (a, c) and right (b, d) show the results obtained by analyzing the HXR images synthesized by the CLEAN and PIXON algorithms, respectively. The panels at the top (a, b) show the measured distances (in arc-seconds) from the brightest pixel (black asterisks) or from the ``center-of-mass'' of brightness (red squares) of the HXR sources to the nearest local maxima of the enhanced PVEC regions (horizontal axis), depending on the measurement error, $\sigma_{\text{HXR}}$, of the HXR source position (vertical axis). The straight oblique dashed line shows the line given by the equation $y=x$. The numbers of data points (and percentages) with the measurement error above/below the distances measured are shown on the left/right of this line in the corresponding color. The graphs in the lower panels (c, d) show the histograms of distributions of the corresponding distances and their fittings by the Gaussian using the least squares method (dots). The found expected values ($\mu$) and standard deviations ($\sigma$) with their errors are shown in the graphs with the corresponding colors. The expected values found are also shown by the straight vertical dashed lines. The range of measurement errors for the distances is grayed out.}
\label{dr_fig}
\end{figure}

For each HXR source constructed with CLEAN and PIXON we calculated two distances to the nearest $j_{r}$-region local maximum. The first one, $dr_{m}$, is the distance between the brightest pixel of a HXR source and the closest local maximum of the nearest $j_{r}$-region. The second one, $dr_{c}$, is the distance between the `center-of-mass' of a HXR source brightness and the closest local maximum of the nearest $j_{r}$-region. Usually, the brightest pixel and the `center-of-mass' of a HXR source brightness are slightly (a few arcseconds) offset from each other. An illustration of determination of these distances for one flare is shown in Figure~\ref{251211_flare_fig}. The results of measurement of $dr_{m}$ and $dr_{c}$ for all HXR sources reconstructed with CLEAN and PIXON are shown in Figure~\ref{dr_fig}. 

Figure~\ref{dr_fig}(a, b) shows the scatter plots of $dr_{m}$ and $dr_{c}$ versus the measurement error, $\sigma_{\text{HXR}}$ (see Section~\ref{Ss-combdata}). First, it can be noted that $\sigma_{\text{HXR}} \lesssim 3^{\prime\prime}$. Second, we found that $dr_{m} > \sigma_{\text{HXR}}$ for 111 (63\%) and 121 (65\%) HXR sources constructed with CLEAN and PIXON, respectively. Similar results were obtained for $dr_{c}$:  $dr_{c} > \sigma_{\text{HXR}}$ for 111 (63\%) and 114 (61\%) HXR sources constructed with CLEAN and PIXON, respectively. Thus, less than $40$\% of HXR sources were located in close proximity to local maxima of $j_{r}$-regions, within the errors ($\pm \sigma_{\text{HXR}}$) of determining the HXR source centers. 

Distributions of $dr_{m}$ and $dr_{c}$ are shown in Figure~\ref{dr_fig}(c, d). They have the shape similar to the normal distribution of a random variable, which is cut off to the left at the zero value. We fit these distributions with a Gaussian function using the least squares method. We obtained the following expected values ($\mu$) and standard deviations ($\sigma$): $\mu\left(r_{m}\right)=1.46 \pm 0.13$ arc-seconds, $\sigma \left(r_{m}\right)=1.82 \pm 0.15$ arc-seconds, and $\mu\left(r_{c}\right)=1.15 \pm 0.31$ arc-seconds, $\sigma \left(r_{c}\right)=2.31 \pm 0.31$ arc-seconds for HXR sources constructed with CLEAN, and $\mu\left(r_{m}\right)=1.50 \pm 0.11$ arc-seconds, $\sigma \left(r_{m}\right)=1.55 \pm 0.13$ arc-seconds, and $\mu\left(r_{c}\right)=1.42 \pm 0.11$ arc-seconds, $\sigma \left(r_{c}\right)=1.51 \pm 0.15$ arc-seconds for HXR sources constructed with PIXON. 

The results obtained with CLEAN and PIXON have some differences, but, in general, they are close to each other. Looking at Figure~\ref{dr_fig}, we can conclude that, although the peaks of the obtained distributions of $dr_{m}$ and $dr_{c}$ are within the measurement error range, $\approx 60$\% of the HXR source centers were located further away from the local maxima of enhanced $j_{r}$-regions than the measurement error $\sigma_{\text{HXR}}$.  


\subsection{DISTRIBUTIONS OF PARAMETERS OF HXR SOURCES AND PVEC UNDER THEM}
\label{Ss-ParamDistr}

\begin{figure}[!t]
\centering
\centerline{\includegraphics[width=0.95\linewidth]{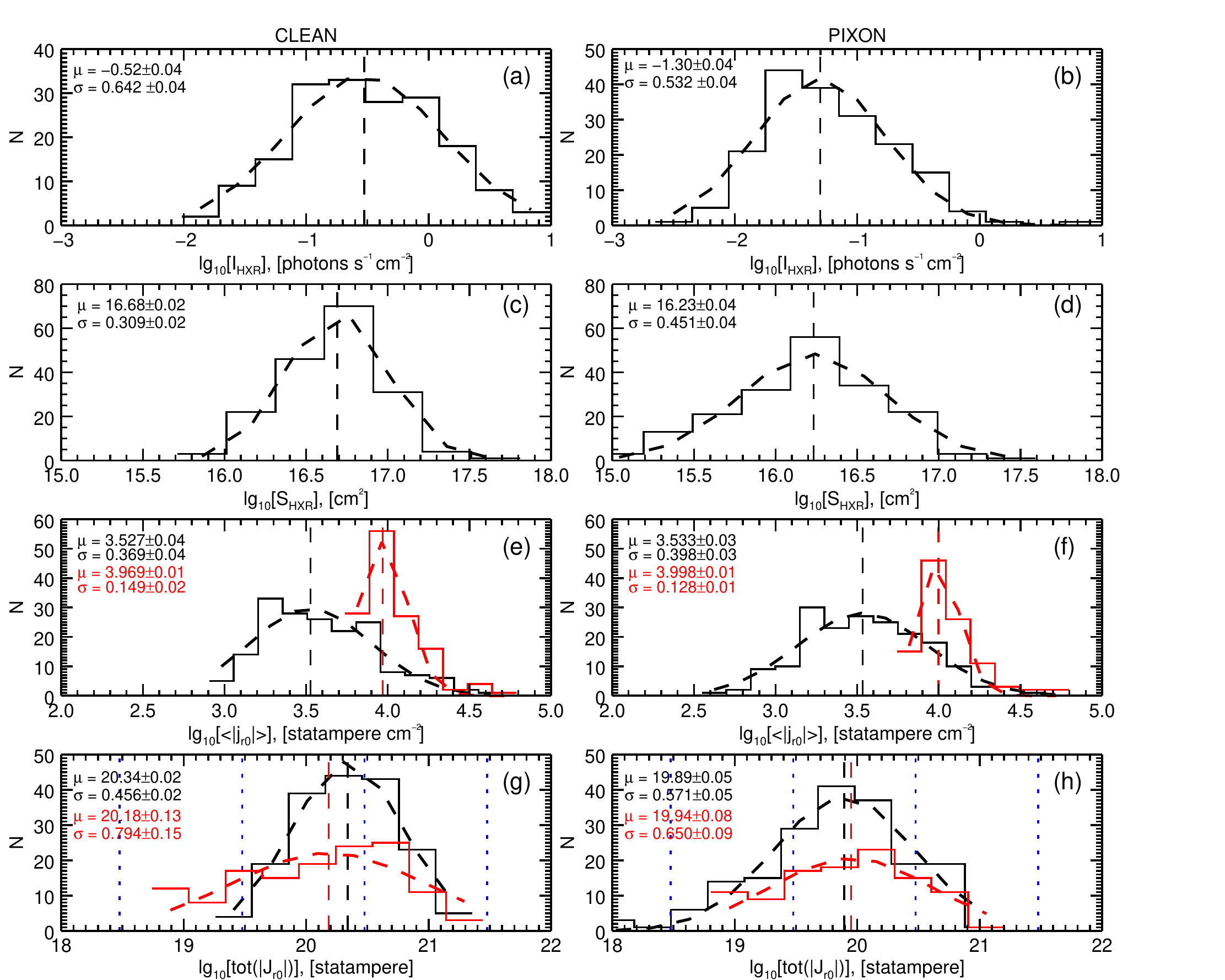}}
\caption{Distributions of various parameters of all the studied HXR (50--100 keV) sources in the selected 48 solar flares. The panels on the left and right show the results obtained by analyzing the HXR images synthesized by the CLEAN and PIXON algorithms, respectively. Each HXR source is defined as a cluster of pixels with an intensity of not less than 90\% of the brightest pixel of a given cluster. Distributions of the parameters obtained are presented by the solid histograms. Their approximations by the Gauss function using the least squares method are shown by the dashed curves of corresponding colors. The found expected values ($\mu$) and standard deviations ($\sigma$) of the gaussians with their errors are shown in the top left corner with the corresponding colors. The expected values are also shown by the straight vertical dashed lines. Distributions of the decimal logarithm of the HXR source intensity ($I_{\text{HXR}}$) and area ($S_{\text{HXR}}$) are shown on (a, b) and (c, d), respectively. Distributions of the decimal logarithm of the mean absolute value of pre-flare PVEC current density $\left\langle \left|j_{r0}\right|\right\rangle$ and total absolute value of pre-flare PVEC $\left|J_{r0}\right|$ under the HXR sources are shown on (e, f) and (g, h), respectively. Black and red curves on (e--h) show the distributions calculated using the ``original'' $j_{r}$-maps and $j_{r}$-maps with values above three standard deviations of the background, respectively. Values equal to $10^{9}$, $10^{10}$, $10^{11}$, and $10^{12}$ A are indicated on (g, h) by blue vertical dotted lines, for convenience.}
\label{4_2_histos_fig}
\end{figure}

The distributions of the decimal logarithms of the flux ($I_{\text{HXR}}$) and area ($S_{\text{HXR}}$) of all HXR sources are presented in Figure~\ref{4_2_histos_fig}(a, b) and Figure~\ref{4_2_histos_fig}(c, d), respectively. We fit them with a Gaussian function using the least squares method. As a result of fitting, the following expected values ($\mu$) and standard deviations ($\sigma$) were obtained for $I_{\text{HXR}}$: $\mu \left( I_{\text{HXR}} \right) = 10^{-0.52 \pm 0.04}$, $\sigma \left( I_{\text{HXR}} \right) = 10^{0.64 \pm 0.04}$ photons~s$^{-1}$~cm$^{-2}$ for CLEAN and $\mu \left( I_{\text{HXR}} \right) = 10^{-1.30 \pm 0.04}$, $\sigma \left( I_{\text{HXR}} \right) = 10^{0.53 \pm 0.04}$ photons~s$^{-1}$~cm$^{-2}$ for PIXON; and for $S_{\text{HXR}}$: $\mu \left( S_{\text{HXR}} \right) = 10^{16.68 \pm 0.02}$, $\sigma \left( S_{\text{HXR}} \right) = 10^{0.31 \pm 0.02}$ cm$^{2}$ for CLEAN and $\mu \left( S_{\text{HXR}} \right) = 10^{16.23 \pm 0.04}$, $\sigma \left( I_{\text{HXR}} \right) = 10^{0.45 \pm 0.04}$ cm$^{2}$ for PIXON. In average, the HXR sources constructed with CLEAN are more intense ($\approx 6$ times) and have larger area ($\approx 2.8$ times), than the HXR sources synthesized with PIXON. 

The distributions of the decimal logarithms of the pre-flare mean absolute value of PVEC density $\left\langle \left|j_{r}\right|\right\rangle$ and total absolute value of PVEC $\left|J_{r}\right|=\sum \left|j_{r} \times \Delta S_{p} \right|$ under the HXR sources are presented in Figure~\ref{4_2_histos_fig}(e, f) and Figure~\ref{4_2_histos_fig}(g, h), respectively. The distributions were separately built for the ``original'' and ``cleaned'' $j_{r}$-maps for comparison. One can see from Figure~\ref{4_2_histos_fig}(e, f) that the distribution of $\left\langle \left|j_{r}\right|\right\rangle$ built with the ``cleaned'' maps is cut off on the left, more narrow, and has higher peak value than the distribution built with the ``original'' maps. The distribution of $\left|J_{r}\right|$ built for the ``cleaned'' maps is wider than the distribution of $\left|J_{r}\right|$ built with the ``original'' maps (Figure~\ref{4_2_histos_fig}(g, h)). This is natural, since the low values (below $3\sigma\left(j_{r}\right)$) in some pixels were zeroed artificially on the ``cleaned'' maps. 

We fit all distributions of $\left\langle \left|j_{r}\right|\right\rangle$ and $\left|J_{r}\right|$ with a Gaussian function using the least squares method. The found expected values and standard deviations are shown in Figure~\ref{4_2_histos_fig}(e--h) and we will not present them here in the text. We just note that the peaks of the $\left\langle \left|j_{r}\right|\right\rangle$ distributions constructed from the ``original'' and ``cleaned'' maps lies in the vicinity of $\approx 10^{3.5}$ and $\approx 10^{4}$ statampere~cm$^{-2}$, respectively. The peaks of the $\left|J_{r}\right|$ distributions constructed from the ``original'' and ``cleaned'' maps are similar and lies in the vicinity of $\approx 10^{20}$ statampere ($\approx 1/3 \times 10^{11}$ A). The peak of $\left|J_{r}\right|$ distribution built for the HXR sources constructed with CLEAN is a bit higher than the one built for the HXR sources constructed with PIXON. This is because the area of the HXR sources constructed with CLEAN is larger, in average, than the area of the HXR sources constructed with PIXON (see above). 176 (99\%) and 148 (79\%) HXR sources reconstructed with CLEAN and PIXON, respectively, have $\left|J_{r}\right| \geq 3 \times 10^{19}$ statampere (or $\geq 10^{10}$ A) calculated using ``original'' maps. However, these numbers decrease to 115 (65\%) and 86 (46\%), respectively, when the calculations were made using ``cleaned'' maps.

\subsection{CORRELATIONS BETWEEN INTENSITY OF HXR SOURCES AND PVEC}
\label{Ss-Corr}

We searched for possible correlations between intensity of the HXR sources and PVEC under them. This was done for several different pairs of parameters related to the HXR source intensity and pre-flare and post-flare PVECs (see section~\ref{Ss-magdata} for the definition of what we call ``pre-flare'' and ``post-flare''). 

\begin{figure}[!t]
\centering
\centerline{\includegraphics[width=0.9\linewidth]{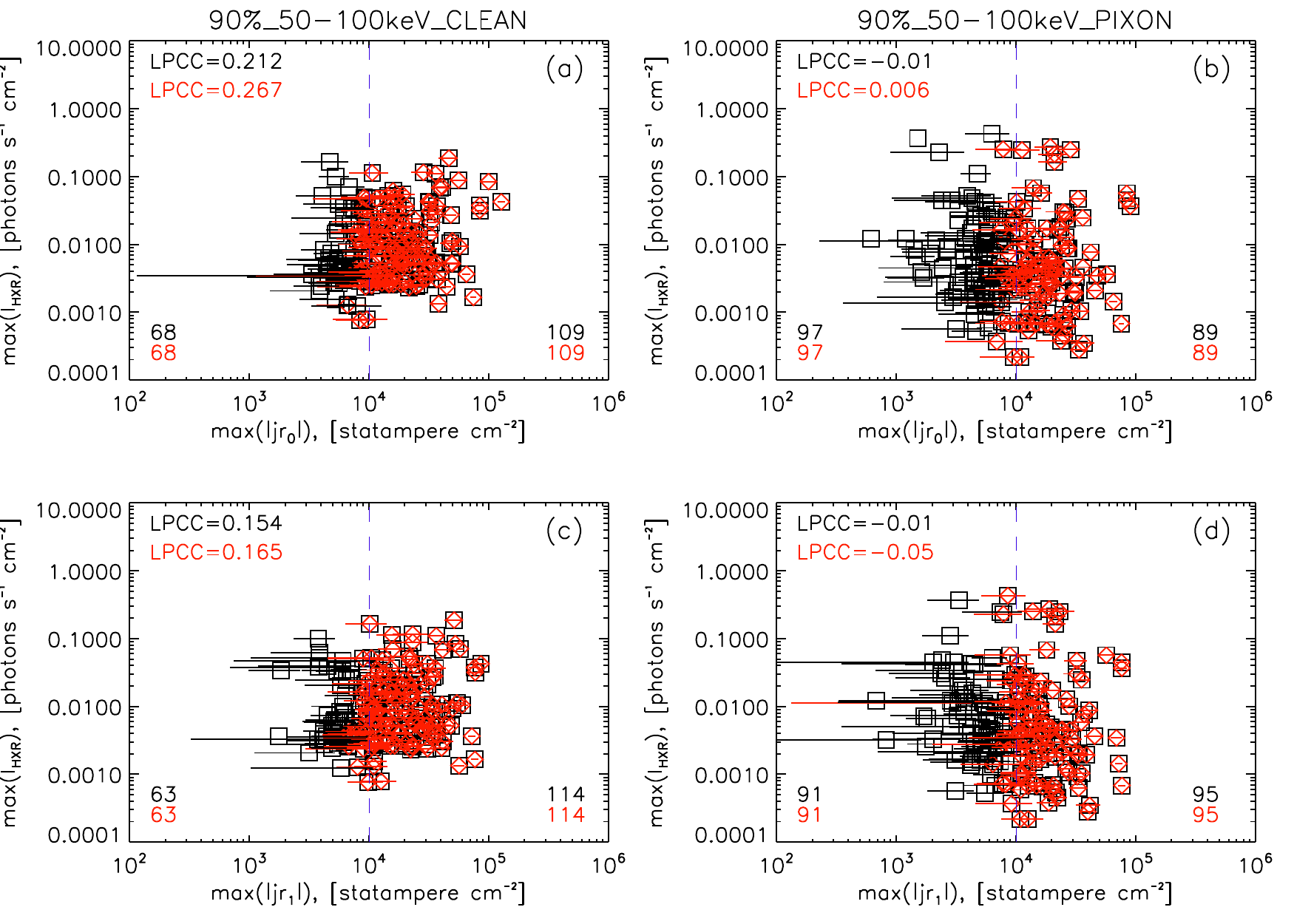}}
\caption{Scatter plots of maximum intensity of the flare HXR (50--100 keV) sources versus maximum absolute value of PVEC $\left|j_{r}\right|$ under the HXR sources. The panels on the left (a, c) and right (b, d) show the results obtained by analyzing the HXR images synthesized by the CLEAN and PIXON algorithms, respectively. The panels at the top (a, b) and bottom (c, d) show the PVEC values $\left|j_{r0}\right|$ and $\left|j_{r1}\right|$ taken before and after the flare impulsive phase, respectively. Black squares and red diamonds show $\left|j_{r}\right|$ values calculated using the `original' $j_{r}$-maps and `cleaned' $j_{r}$-maps with values above three standard deviations of the background, respectively. The estimated errors are shown by horizontal and vertical bars. The values of the linear Pearson correlation coefficient (LPCC) are shown in the upper left corner in appropriate color. The threshold level $j_{r}^{thr}$ is shown by the blue dashed vertical line. The numbers of HXR sources with $\left|j_{r}\right| < j_{r}^{thr}$ and $\left|j_{r}\right| \geq j_{r}^{thr}$ are shown in the left and right bottom corners in appropriate color, respectively.}
\label{max_abs_jr_ihxrc_fig}
\end{figure}

Figure~\ref{max_abs_jr_ihxrc_fig} shows the scatter plots of the maximum intensity of the HXR sources versus the maximum of pre-flare $\left|j_{r0}\right|$ or post-flare $\left|j_{r1}\right|$ under the HXR sources constructed with the CLEAN and PIXON algorithms. The values of the linear Pearson correlation coefficient (LPCC) are shown in the figures. Based on the shape of the clouds of data points and the LPCC values obtained, we conclude that there are no obvious correlations between the considered parameters.

\begin{figure}[!t]
\centering
\centerline{\includegraphics[width=0.9\linewidth]{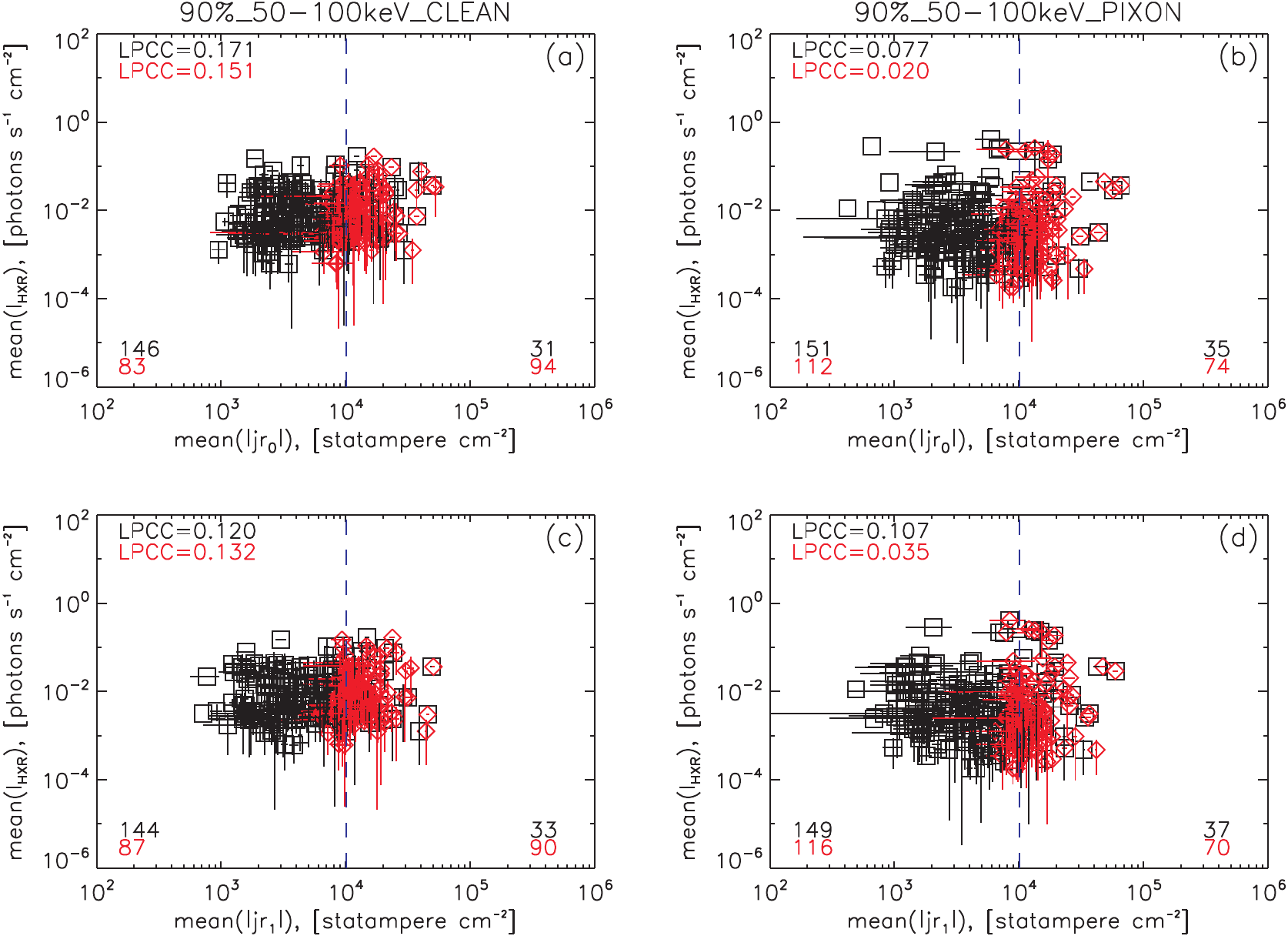}}
\caption{The same as in figure~\ref{max_abs_jr_ihxrc_fig}, except that the mean values are presented instead of the maximum ones.}
\label{mean_abs_jr_ihxrc_fig}
\end{figure}

Figure~\ref{mean_abs_jr_ihxrc_fig} shows the scatter plots of the average intensity of the HXR sources versus the average pre-flare $\left|j_{r0}\right|$ or post-flare $\left|j_{r1}\right|$ under the HXR sources constructed with the CLEAN and PIXON algorithms. Based on the shape of the clouds of data points and the LPCC values obtained, we make similar conclusion that there are no correlations between the considered parameters.

\begin{figure}[!t]
\centering
\centerline{\includegraphics[width=0.9\linewidth]{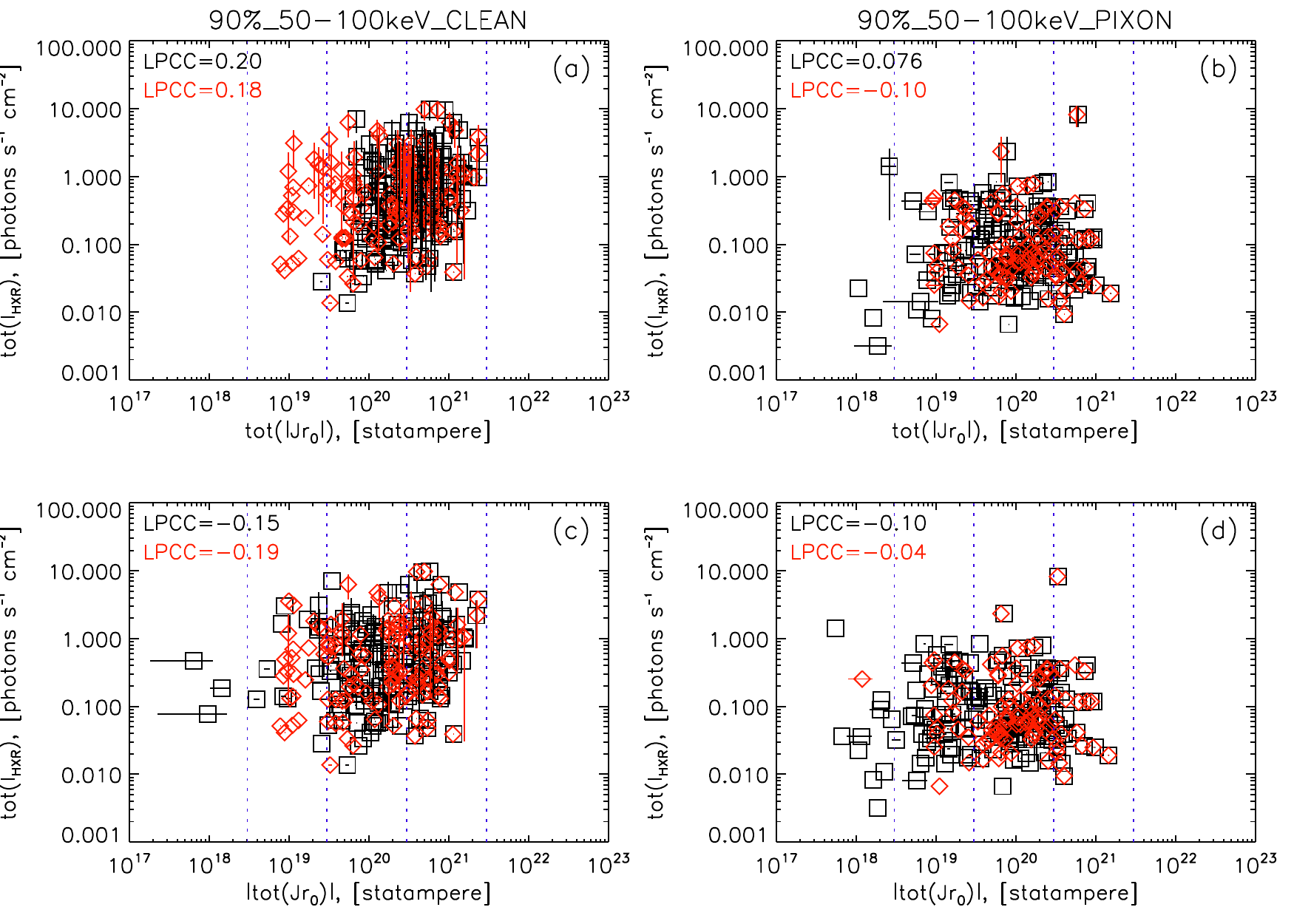}}
\caption{Scatter plots of total intensity of the flare HXR (50--100 keV) sources (within 90\% of maximum brightness) versus total absolute value of pre-flare PVEC $\left|J_{r0}\right|$ (a, b) or absolute value of total $J_{r0}$ (c, d) under the HXR sources. The panels on the left (a, c) and right (b, d) show the results obtained by analyzing the HXR images synthesized by the CLEAN and PIXON algorithms, respectively. Black squares and red diamonds show values calculated using the `original' $j_{r}$-maps and `cleaned' $j_{r}$-maps with values above three standard deviations of the background, respectively. The estimated errors are shown by horizontal and vertical bars. The values of the linear Pearson correlation coefficient (LPCC) are shown in the upper left corner in appropriate color. Values of $10^{9}$, $10^{10}$, $10^{11}$, and $10^{12}$ A are indicated by blue vertical dotted lines (for convenience).}
\label{tot_abs_jr_ihxrc_fig}
\end{figure}

Figure~\ref{tot_abs_jr_ihxrc_fig} shows the scatter plots of the total intensity of the HXR sources versus the total absolute value of pre-flare PVEC $\left|J_{r0}\right| = \sum \left| j_{r} \times \Delta S_{p} \right|$ (panels (a, b)) or absolute value of the total pre-flare PVEC $\left|J^{\prime}_{r0}\right|= \left| \sum \left( j_{r} \times \Delta S_{p} \right) \right|$ (panels (c, d)) under the HXR sources. We see no correlations between the considered parameters.

\begin{figure}[!t]
\centering
\centerline{\includegraphics[width=0.9\linewidth]{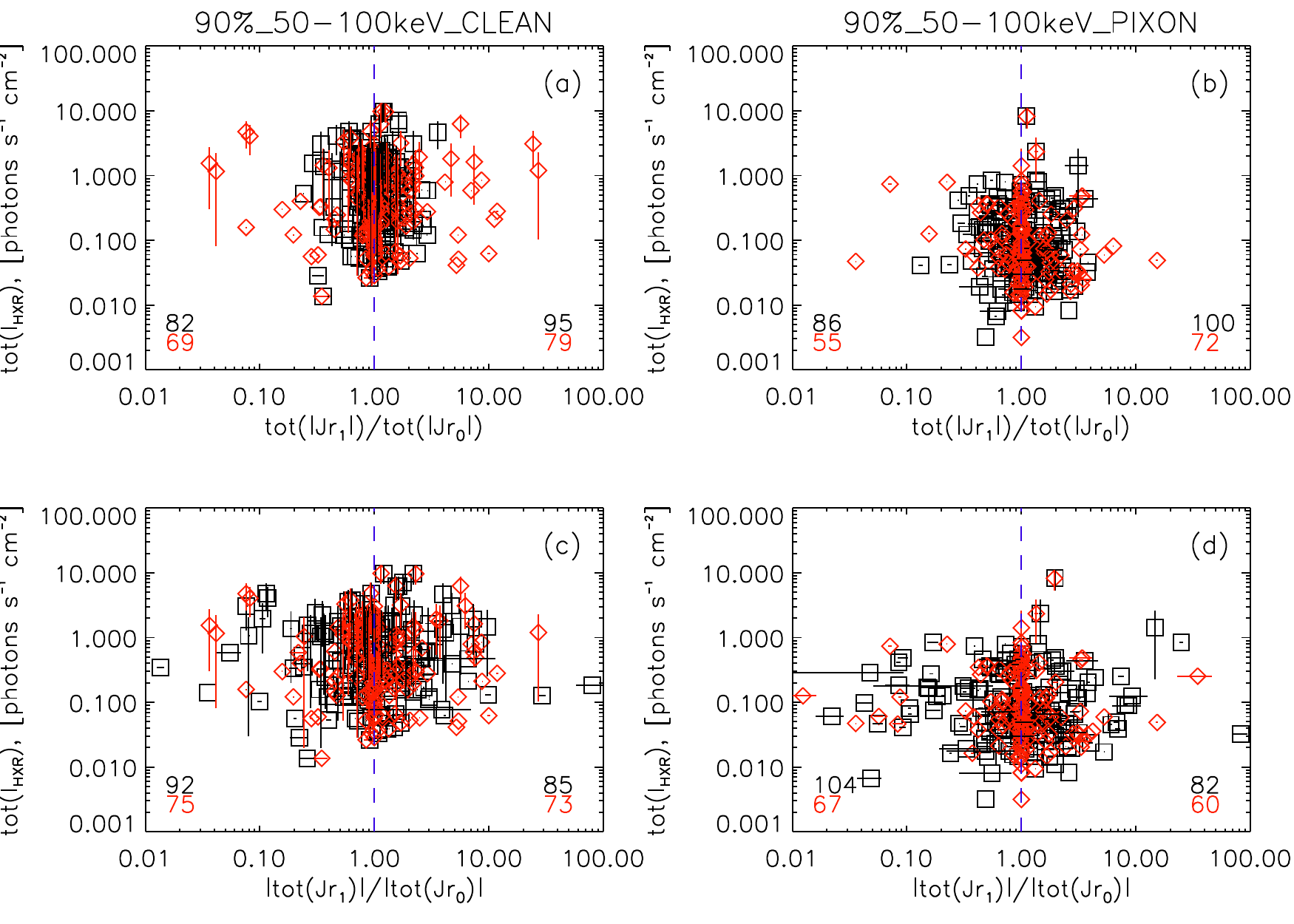}}
\caption{Scatter plots of total intensity of the flare HXR (50--100 keV) sources (within 90\% of maximum brightness) versus ratio of total absolute PVEC values (a, b) or absolute total PVEC values (c, d) after ($J_{r1}$) and before ($J_{r0}$) the flare impulsive phase under the HXR sources. The panels on the left (a, c) and right (b, d) show the results obtained by analyzing the HXR images synthesized by the CLEAN and PIXON algorithms, respectively. Black squares and red diamonds show values calculated using the `original' $j_{r}$-maps and `cleaned' $j_{r}$-maps with values above three standard deviations of the background, respectively. The estimated errors are shown by horizontal and vertical bars. The numbers of data points with values lower and higher than unity (marked by the blue vertical dashed line) are shown in the lower left and right corners with the corresponding color, respectively.}
\label{ratio_tot_jr_ihxrc_fig}
\end{figure}

Finally, Figure~\ref{ratio_tot_jr_ihxrc_fig} shows the scatter plots of the total intensity of the HXR sources versus the ratio of the total absolute PVECs after and before the flare impulsive phase, $\left|J_{r1}\right| / \left|J_{r0}\right|$ (panels (a, b)), or the ratio of the absolute total PVECs after and before the flare impulsive phase, $\left|J^{\prime}_{r1}\right| / \left|J^{\prime}_{r0}\right|$ (panels (c, d)) under the HXR sources. This could show whether intensity of the HXR sources was related to changes of the total PVECs under the HXR sources or not. We did not calculate LPCC between these parameters for obvious reason. The values of these ratios lie in a cloud of dots around unity. We calculated numbers of HXR sources with $\left|J_{r1}\right| / \left|J_{r0}\right| < 1$ or $\left|J_{r1}\right| / \left|J_{r0}\right| > 1$ and also with $\left|J^{\prime}_{r1}\right| / \left|J^{\prime}_{r0}\right| < 1$ or $\left|J^{\prime}_{r1}\right| / \left|J^{\prime}_{r0}\right| > 1$. They are shown in opposite lower corners in the corresponding panels of Figure~\ref{ratio_tot_jr_ihxrc_fig}. For the values obtained using both the ``original'' and ``cleaned'' $j_{r}$-maps, there is no systematic difference between PVECs after and before the flares. The numbers of data points (i.e. HXR sources) with values greater than and less than one are approximately equal. Scatter plots for the mean intensity of the HXR sources versus the ratios $\left| \left\langle j_{r1} \right\rangle \right| / \left| \left\langle j_{r0} \right\rangle \right|$ and $\left\langle \left|  j_{r1}  \right| \right\rangle / \left\langle \left|  j_{r0}  \right| \right\rangle$ look similar. We do not present them in order not to increase the volume of the article. 

We also computed Spearman's and Kendall's rank correlation coefficients. The results are the same --- no correlation found.

To summarize briefly, we did not find evidences of correlation between intensity of the HXR sources and PVECs under them, as well as PVEC changes during the flares.

\subsection{MAGNETIC FIELDS AND ITS CHANGES UNDER HXR SOURCES}
\label{Ss-magHXR}

\begin{figure}[!t]
\centering
\centerline{\includegraphics[width=0.9\linewidth]{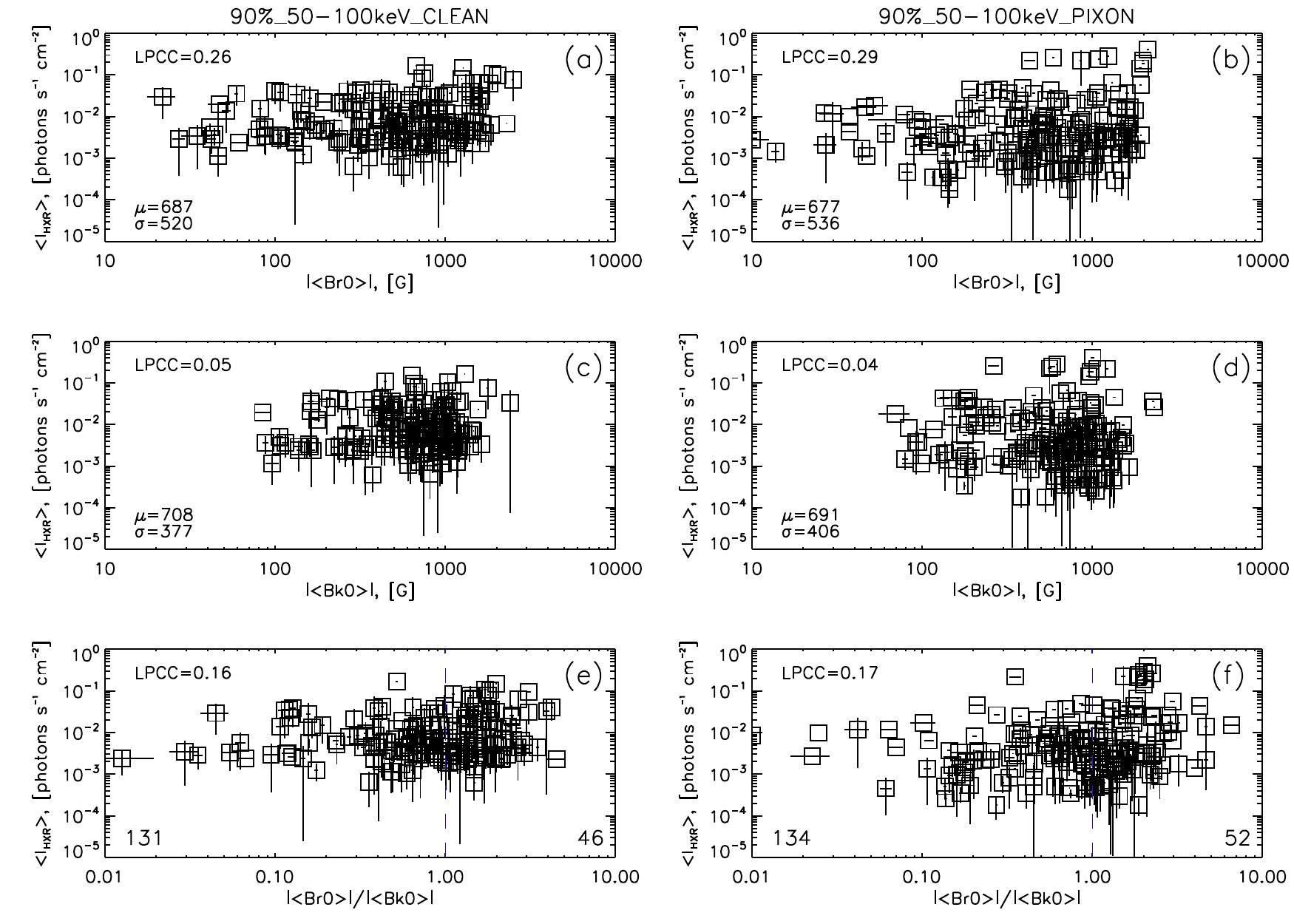}}
\caption{Scatter plots of mean intensity of the flare HXR (50--100 keV) sources versus absolute values of the pre-flare radial magnetic component $B_{r0}$ (a, b), tangential magnetic component $B_{k0}$ (c, d), and their ratio. The panels on the left (a, c, e) and right (b, d, f) show the results obtained by analyzing the HXR images synthesized by the CLEAN and PIXON algorithms, respectively. The estimated errors are shown by horizontal and vertical bars. The values of the linear Pearson correlation coefficient (LPCC) are shown in the upper left corners. The found expected values ($\mu$) and standard deviations ($\sigma$) are shown in the bottom left corners on (a--d). The numbers of HXR sources with $\left|\left\langle B_{r0}\right\rangle\right| < \left|\left\langle B_{k0}\right\rangle\right|$ and $\left|\left\langle B_{r0}\right\rangle\right| \geq \left|\left\langle B_{k0}\right\rangle\right|$ are shown in the left and right bottom corners of (e, f), respectively.}
\label{mean_ihxr_br_bk_fig}
\end{figure}

It is also interesting to check for possible correlations between intensity of the HXR sources and different components of the photospheric magnetic field under them. For this, we made the scatter plots of the average intensity of the HXR sources, $\left\langle I_{\text{HXR}}\right\rangle$, versus the absolute average values of the pre-flare radial magnetic field under the HXR sources, $\left| \left\langle B_{r0} \right\rangle \right|$ (Figure~\ref{mean_ihxr_br_bk_fig}(a, b)), tangential field, $\left| \left\langle B_{k0} \right\rangle \right|$ (Figure~\ref{mean_ihxr_br_bk_fig}(c, d)), and their ratio $\left| \left\langle B_{r0} \right\rangle \right| / \left| \left\langle B_{k0} \right\rangle \right|$ (Figure~\ref{mean_ihxr_br_bk_fig}(e, f)). The values of the linear Pearson correlation coefficient (LPCC) are shown in the figures. Based on the LPCC values obtained, we conclude that there are no significant correlations between the considered parameters. We also computed Spearman's and Kendall's rank correlation coefficients, which also did not show correlation between the parameters considered. The expected values ($\mu$) and standard deviations ($\sigma$) of the radial and tangential magnetic field components are shown in the figures. On average, the tangent magnetic component under the HXR sources is slightly higher than the radial component. This is also visible in Figure~\ref{mean_ihxr_br_bk_fig}(e, f), which shows the ratio of the components. The number of HXR sources in which the tangential component is higher than the radial component, i.e. where $\left| \left\langle B_{r0} \right\rangle \right| / \left| \left\langle B_{k0} \right\rangle \right| < 1$, is approximately three times larger (the corresponding numbers are shown in the bottom left and right corners in Figure~\ref{mean_ihxr_br_bk_fig}(e, f)).

\begin{figure}[!t]
\centering
\centerline{\includegraphics[width=0.9\linewidth]{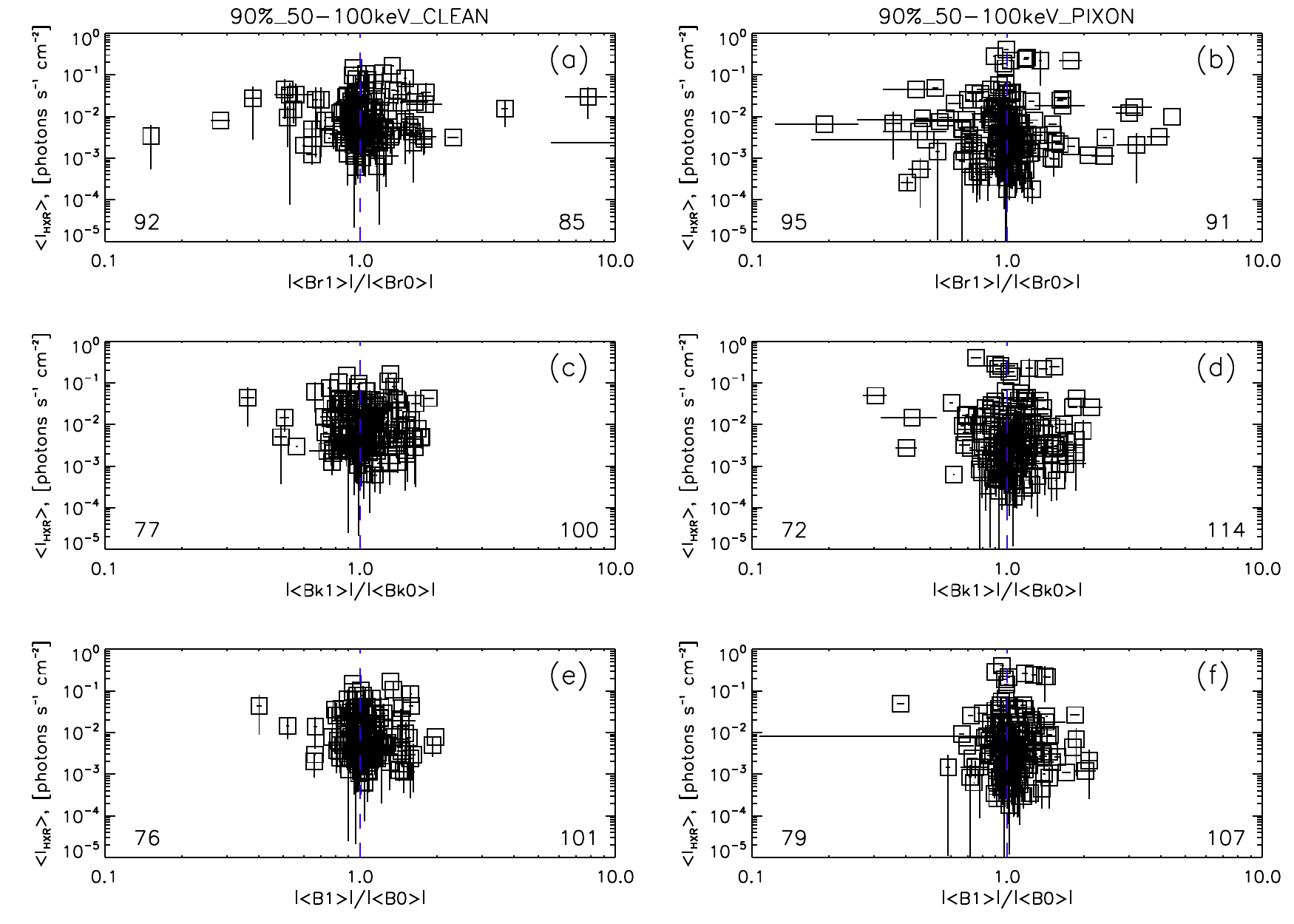}}
\caption{Scatter plots of mean intensity of the flare HXR (50--100 keV) sources versus ratios of absolute values of post-flare and pre-flare radial magnetic components, $\left| \left\langle B_{r1} \right\rangle \right| / \left| \left\langle B_{r0} \right\rangle \right|$ (a, b), tangential magnetic components, $\left| \left\langle B_{k1} \right\rangle \right| / \left| \left\langle B_{k0} \right\rangle \right|$ (c, d), and full magnetic field vectors, $\left| \left\langle B_{1} \right\rangle \right| / \left| \left\langle B_{0} \right\rangle \right|$ (e, f). The panels on the left (a, c, e) and right (b, d, f) show the results obtained by analyzing the HXR images synthesized by the CLEAN and PIXON algorithms, respectively. The estimated errors are shown by horizontal and vertical bars. The numbers of HXR sources with magnetic field ratios lower and higher than one are shown in the left and right bottom corners, respectively.}
\label{mean_ihxr_ratio_br_bk_b_fig}
\end{figure}

We also checked for possible correlations between intensity of the HXR sources and changes of magnetic fields under them during the flares. For this, we built the scatter plots of the average intensity of the HXR sources, $\left\langle I_{\text{HXR}}\right\rangle$, versus the ratios of the absolute average values of the post-flare and pre-flare radial magnetic components, $\left| \left\langle B_{r1} \right\rangle \right| / \left| \left\langle B_{r0} \right\rangle \right|$ (Figure~\ref{mean_ihxr_ratio_br_bk_b_fig}(a, b)), tangential magnetic components, $\left| \left\langle B_{k1} \right\rangle \right| / \left| \left\langle B_{k0} \right\rangle \right|$ (Figure~\ref{mean_ihxr_ratio_br_bk_b_fig}(c, d)), and full magnetic vectors, $\left| \left\langle B_{1} \right\rangle \right| / \left| \left\langle B_{0} \right\rangle \right|$ (Figure~\ref{mean_ihxr_ratio_br_bk_b_fig}(e, f)), under the HXR sources. There is no evident dependency of intensity of the HXR sources from the changes of the radial magnetic component. Data points lie approximately evenly around a value of 1. For the tangential component, the situation is generally similar. However, it can be noted that the number of data points (i.e. HXR sources), at which the magnitude of the post-flare tangential component has increased, is a bit ($\approx 20$\%) larger than the number of data points with a decreased tangential component. Almost the same picture is for the absolute value of magnetic field vector. This is not surprising, since in most HXR sources the tangential component exceeds the radial one (see above). An increase in the tangential component of the photospheric magnetic field has been noted for some flares \citep[e.g., ][]{Wang1994,Sun2012,Petrie2013,Sharykin2019}. 

To summarize, we did not find obvious correlation between intensity of the HXR sources and photospheric magnetic field under them, and also their changes during the impulsive phase of the flares studied.  

\subsection{ASYMMETRIES OF INTENSITY, MAGNETIC FIELDS AND PVEC OF PAIRED HXR SOURCES}
\label{Ss-asymm}

Finally, we conducted an analysis of a special sub-sample of HXR sources, namely the paired HXR sources. By such sources we mean a pair of simultaneously observed HXR sources, $\text{HXR}_{1}$ and $\text{HXR}_{2}$, for which the average values of the pre-flare radial magnetic field, $\left\langle B_{r0} \right\rangle$, and PVEC density, $\left\langle j_{r0} \right\rangle$, have opposite signs, i.e. for which $\left\langle B^{\text{HXR}_{1}}_{r0} \right\rangle \left\langle B^{\text{HXR}_{2}}_{r0} \right\rangle < 0$ and $\left\langle j^{\text{HXR}_{1}}_{r0} \right\rangle \left\langle j^{\text{HXR}_{2}}_{r0} \right\rangle < 0$. Such sources can be considered, with a high probability, as the conjugate footpoints of current-carrying flare loops. They are indicated by the subscript `p' in columns 7 and 8 in Table~\ref{tab:flareinfo}. The sub-samples of such HXR sources turned out to be small: only 24 and 12 pairs for the CLEAN and PIXON algorithms, respectively. 

The scatter plots of intensity of the paired HXR sources versus PVEC and magnetic field parameters look similar (except fewer data points) to the scatter plots built for the complete set of all HXR sources shown in the previous sections. For brevity, we will not show them here. We did not find significant correlation between these parameters, as above. 

\begin{figure}[!t]
\centering
\centerline{\includegraphics[width=0.9\linewidth]{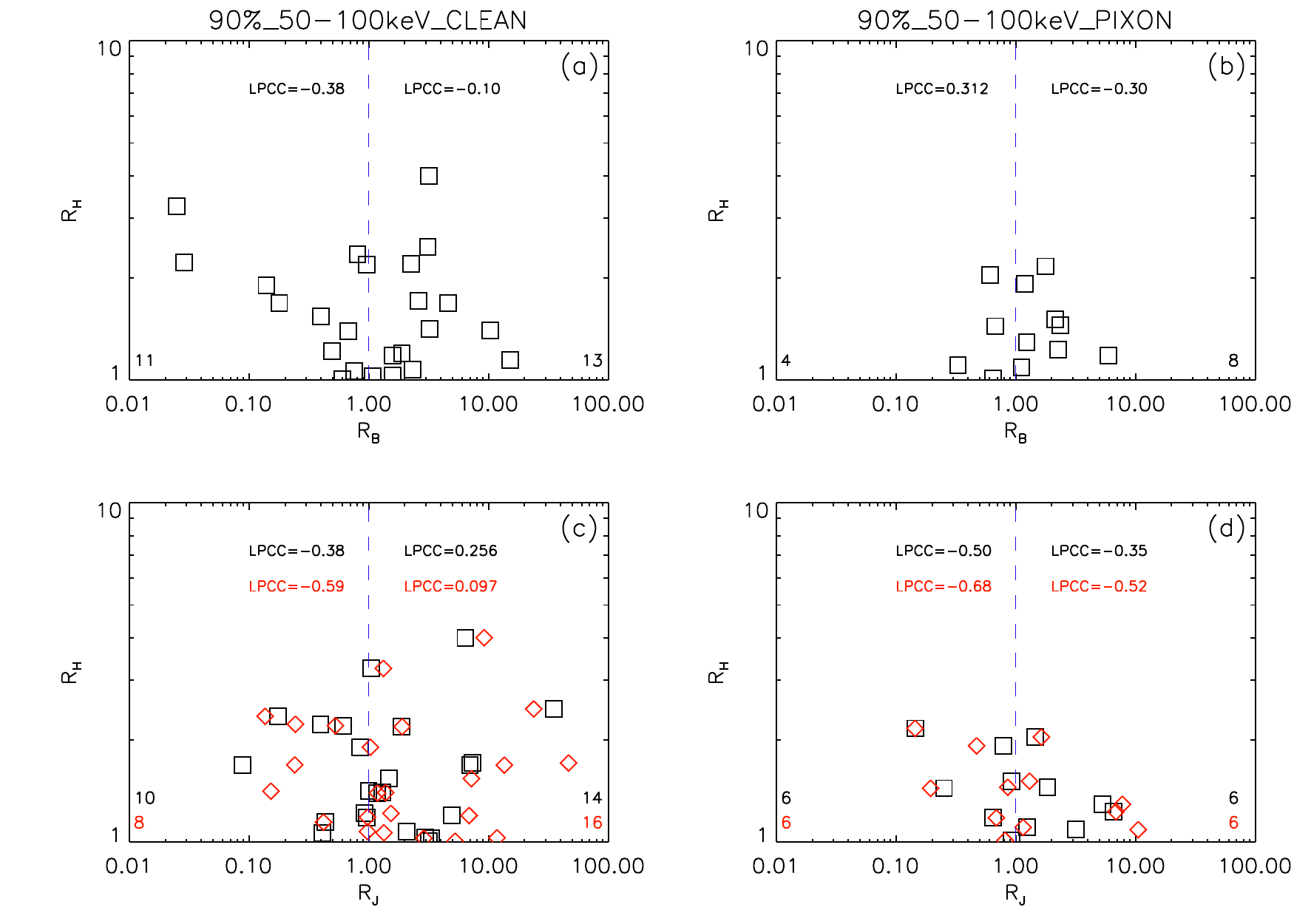}}
\caption{Scatter plots of the 50--100 HXR source total flux ratio, $R_{H}$, versus the absolute mean radial magnetic field ratio, $R_{B}$ (a, b), and the absolute mean PVEC ratio, $R_{J}$ (c, d), under the HXR sources. The panels on the left (a, c) and right (b, d) show the results for the HXR sources constructed with the CLEAN and PIXON algorithms, respectively. Black squares and red diamonds show values calculated using the `original' and `cleaned' pre-flare $J_{r0}$-maps, respectively. The numbers of data points with $R_{B}$ or $R_{J}$ less and greater than unity are shown in the bottom left and right corners, respectively. The corresponding values of the linear Pearson correlation coefficient (LPCC) are shown left and right of the dashed vertical line denoting the value 1.}
\label{rh_rb_rj_fig}
\end{figure}

Here we consider the flux ratio of the paired HXR sources, $R_{H}=I^{\text{HXR}_{2}}/I^{\text{HXR}_{1}}$, where we notated HXR sources such as $I^{\text{HXR}_{2}} \geq I^{\text{HXR}_{1}}$ (i.e. $R_{H} \geq 1$), and compare it with the absolute value of the pre-flare mean radial magnetic field ratio $R_{B}=\left| \left\langle  B_{r0}^{\text{HXR}_{2}} \right\rangle / \left\langle B_{r0}^{\text{HXR}_{1}} \right\rangle \right|$, and also with the absolute value of the pre-flare PVEC ratio, $R_{J}=\left| J_{r0}^{\text{HXR}_{2}} / J_{r0}^{\text{HXR}_{1}} \right|$, calculated for area under the HXR sources. The scatter plots of these ratios are shown in Figure~\ref{rh_rb_rj_fig}, again for the HXR sources constructed with the CLEAN and PIXON algorithms, for comparison. The values of $R_{H}$, $R_{B}$, and $R_{J}$ are in the ranges $1.006 \leq R_{H} \leq 4.006$, $0.025 \leq R_{B} < 15.202$, $0.088 \leq R_{J} < 46.688$, respectively. The asymmetry of fluxes of the paired HXR sources is, in general, much less than the asymmetry of the magnetic fields and PVECs under them.

On the scatter plot of $R_{H}$ versus $R_{B}$ made using CLEAN (Figure~\ref{rh_rb_rj_fig}(a)), two sets of data points can be distinguished: (1) with $R_{B} < 1$ and (2) with $R_{B} > 1$. There is a weak anti-correlation between $R_{H}$ and $R_{B}$ for $R_{B} < 1$. The linear Pearson correlation coefficient, $\text{LPCC}=-0.38$, is shown in the figure. This may indicate the presence of an asymmetric magnetic mirror effect in these events, when stronger HXR sources are located in weaker magnetic field \citep[e.g.][]{Li1997,Yang2012}. However, the small number of data points and weak anti-correlation, observed for the HXR sources constructed with CLEAN only, do not allow seriously discussing this effect here.

Approximately similar situation is for the ratios $R_{H}$ and $R_{J}$. There is a weak anti-correlation between $R_{H}$ and $R_{J}$ for the HXR sources with $R_{J} < 1$. It is found for the HXR sources constructed both with CLEAN (Figure~\ref{rh_rb_rj_fig}(c)) and PIXON (Figure~\ref{rh_rb_rj_fig}(d)). This may indicate that some stronger HXR sources could be located in weaker PVEC regions. There is also a weak anti-correlation between $R_{H}$ and $R_{J}$ for the HXR sources constructed only with PIXON, for which $R_{J} > 1$. As above, we think it is too prematurely to discuss these effects seriously because of the small number of data points and the weak value of anti-correlation.    

\section{SUMMARY OF RESULTS AND DISCUSSION}
\label{S-Discussion}

In this section we will summarize and discuss the results of the data analysis performed. We found the presence of multiple enhanced $j_{r}$-regions ($\left| j_{r} \right| \geq j^{thr}_{r} \approx 10.1 \times 10^{3}$ statampere~cm$^{-2}$) in the parent active regions of all 48 solar flares studied. The choice of this specific value of $j^{thr}_{r}$ is determined by the shape of $j_{r}$-distributions obtained with the \texttt{SHARP} data series. As it was shown in \citet{Zimovets2019}, the $j_{r}$-distributions below $j^{thr}_{r}$ have a Gaussian shape and can be composed mainly from data noise. Above this threshold value, the $j_{r}$-distributions have a power-law shape, which can be physically meaningful, representing some real processes of formation and distribution of PVECs in active regions. Different threshold values were used in different works studying relationship between PVECs and flare emission sources: e.g., \citet{Moreton1968} used $j^{thr}_{r} \approx 2.5 \times 10^{3}$ statampere~cm$^{-2}$, \citet{Li1997} used $0.9 \times 10^{3} \leq j^{thr}_{r} = 3 \sigma\left(j_{r}\right) \leq 3.2 \times 10^{3}$ statampere~cm$^{-2}$ for different events, \citet{Musset2015} used $j^{thr}_{r} = 3 \times 10^{4}$ statampere~cm$^{-2}$. Some of them are higher or lower than the threshold level we used. However, the data of different instruments were used in different studies, and not all of them clearly explained the choice of the threshold value used, unlike in the present work. 
 
The enhanced $j_{r}$-regions mainly have a shape of numerous $j_{r}$-islands with a characteristic size of several arc-seconds or less numerous elongated $j_{r}$-ribbons up to several tens of arc-seconds in length. Despite the fact that $j_{r}$-islands are more numerous, HXR sources overlapped with $j_{r}$-ribbons about 1.5 times more often. For 48 solar flares studied we reconstructed HXR (50--100 keV) sources in 81 time intervals, corresponding to the main HXR peaks, and found 177 and 186 HXR sources in the images synthesized with the CLEAN and PIXON algorithms, respectively.

We found that $\approx 70$\% of all HXR sources overlapped, at least partially, with one or a few enhanced $j_{r}$-regions, within the estimated source center error, $\sigma^{\text{HXR}} \approx \pm 3^{\prime\prime}$. This is close to the results obtained by \citet{Moreton1968} and \citet{Zvereva1970}, who found that $\approx 70-80$\% flare H$_{\alpha}$ knots overlap with enhanced $j_{r}$-regions, although their $j_{r}$-threshold level was four times lower. We found also that 30\%--42\% and 22\%--31\% of the HXR sources overlapped with local and global maxima, respectively, of enhanced $j_{r}$-regions. Only 5\%--9\% of all HXR sources overlapped with the major $j_{r}$-maxima (positive or negative) of an entire parent active region. 

In $\approx 90$\% of the flares studied at least one HXR source was in enhanced $j_{r}$-regions. In 60\%--75\% and 52\%--65\% of the flares at least one HXR source was in local and global, respectively, maxima of enhance $j_{r}$-regions. In 17\%--23\% of the flares at least one HXR source was in the major $j_{r}$-maxima of an entire active region. 

The distribution of the total absolute PVEC values under the HXR sources was approximated with a Gaussian with the expected value $\approx 10^{20}$ statampere (or $\approx 1/3 \times 10^{11}$ A). 176 (99\%) and 148 (79\%) HXR sources reconstructed with CLEAN and PIXON, respectively, have $\left|J_{r}\right| \geq 3 \times 10^{19}$ statampere (or $\geq 10^{10}$ A) calculated using ``original'' maps. These numbers decrease to 115 (65\%) and 86 (46\%), respectively, when the calculations were made using ``cleaned'' maps (with $\left|j_{r}\right| > 3 \sigma\left(j_{r}\right)$). Here we need to note that we determined the HXR sources by the specific way, as clusters of bright pixels with intensity above 90\% of the maximum value in the cluster. Thus, we dealt with the central core of the HXR sources and did not care about their periphery. The total PVEC values under the HXR sources could be several times higher if we would use a lower contour level (e.g., 50\% of maximal) to restrict the HXR sources.   

The results presented above generally indicate a close relationship between the flare HXR sources and PVECs. It may seem that they provide evidence in favor of the current-interruption models in which longitudinal currents play a key role in the process of energy release and acceleration of charged particles \citep[e.g.][]{Alfven1967,Spicer1981,Zaitsev1998,Zaitsev2015,Zaitsev2016a}. According to \citet{Zaitsev2016a}, a powerful flare with effective electron acceleration and HXR radiation can occur in a current-carrying loop when the total current in it exceeds $\sim 3 \times 10^{19}$ statampere. This may be considered as a necessary but not sufficient condition for a powerful flare in the current-interruption model. Below we summarize the results, which, in our opinion, do not fit with this model. 

At first, it should be noted that $\approx 30$\% of all HXR sources were outside enhanced $j_{r}$-regions. In 5 flares ($\approx 10$\%) all found HXR sources did not overlap with the enhanced $j_{r}$-regions. In $\approx 48$\% of the flares, part of the HXR sources did not overlap with the enhanced $j_{r}$-regions. In two such flares, HXR sources overlapped with the enhanced $j_{r}$-regions at some time intervals, while they all were outside the enhanced $j_{r}$-regions during other HXR peaks. Thus, in total, more than in half ($\approx 58$\%) of the flares studied there were HXR sources outside the enhanced $j_{r}$-regions. Only in $\approx 17$\% of the flares studied all HXR sources overlapped with local or global maxima of the enhanced $j_{r}$-regions. In most of these cases, the enhanced $j_{r}$-regions were tiny islands consisting of only 1--4 pixels and having an angular size much smaller than a size of the corresponding HXR sources. Such tiny $j_{r}$-islands may represent not fully `cleaned' data noise. This issue requires further investigation. 

It should be noted here that in the model of Zaitsev and Stepanov \citep{Zaitsev2015,Zaitsev2016a}, the spatial coincidence of PVEC and HXR maxima is not necessary. The process of generating a pulse of longitudinal electric field, $E_{r}$, as a result of the development of the balloon mode of the magnetic Rayleigh-Taylor instability, is considered in \citep{Zaitsev2015}. The instability criterion is given by their formula (13). From this formula it follows that the instability occurs when outer shell of a current-carrying magnetic loop is heated, as well as with a sharp increase in the velocity of the convective plasma flow at the loop foot. The development of the instability does not depend directly on the magnitude of the current in the loop. Therefore, the most effective acceleration of particles (electrons) and the maximum of bremsstrahlung HXR emission do not have to coincide in space with the maximum current density. For example, the maximum current may flow in the central region of the loop, while Rayleigh-Taylor instability and particle acceleration (as well as HXRs) may occur at its periphery, where the current may be weaker. Thus, the observational result that the maxima of the HXR sources are located predominantly on the periphery of the enhanced $j_{r}$-regions cannot serve as a strong evidence that the model under consideration is not valid. However, the established fact that there were HXR sources outside the enhanced $j_{r}$-regions in more than half of the considered flares can serve as an argument against this model.

Secondly, we did not find correlation between intensity of the HXR sources and different characteristics of PVECs under them. We checked correlation between average and maximum intensity of the HXR sources and average and maximum pre-flare and post-flare PVEC density, and also we examined correlation between the total HXR source intensity and pre-flare and post-flare PVEC under the HXR sources (see our definition of ``pre-flare'' and ``post-flare'' in section~\ref{Ss-magdata}). The ratio of post-flare to pre-flare PVECs also did not show significant correlation with intensity of the HXR sources. Additionally, we separately analyzed a sub-sample of conjugate footpoint HXR sources, which may be more consistent with the current-interruption models. We did not find correlation of the same parameters for this sub-sample of paired HXR sources. However, the current-interruption models assume a relationship between a current flowing along flare loops, $J_{r}$, and a magnitude of the longitudinal electric field, $E_{r}$, generated when the longitudinal current is interrupted. In particular, in the model developed by \citet{Zaitsev2016a}, for a sufficiently strong pre-flare electric current, $J_{r0} \geq 3 \times 10^{19}$ statampere, flowing along a pre-flare loop, a generated pulse of $E_{r}$ strongly depends on $J_{r0}$, as $E_{r} \propto J^{3}_{r0}$, and can exceed Dreicer field. In such case, the bulk of electrons in the site of this pulse is accelerated in the runaway mode. Obviously, the flux of the emitted bremsstrahlung HXR radiation should depend in a more complex way on the pre-flare longitudinal current, but in any case some correlation is expected. However, as mentioned above, we were not able to detect correlation between the intensity of the HXR sources and PVECs under them. Also, intuitively, it is expected within these models that the longitudinal current in flare loops should decrease during the impulsive phase, since at least part of its energy should be transformed into the kinetic energy of accelerated particles and heated plasma, as well as electromagnetic radiation and plasma waves. However, we did not find a systematic decrease in PVECs in the region of the studied HXR sources. In more than half of the sources, the ratio of post-flare to pre-flare PVECs was greater than or equal to unity. 

It is difficult to explain these results within the current-interruption flare models. On the other hand, the presence of enhanced $j_{r}$-regions in all the active regions studied and in close proximity (within $\pm 3^{\prime\prime}$) to the majority of the flare HXR sources seems natural. It is known that the magnetic field in the active regions of the Sun, especially in the core of flare regions, deviates from the potential state \citep[e.g.][]{Schrijver2005,Sadykov2014,Schrijver2016}. This means that spatially separated electric currents flow there \citep[e.g., ][]{Fleishman2018}. Since, basically, the plasma beta in the corona in the active regions is less than unity, currents flow mainly along the magnetic field \citep[e.g.][]{Wiegelmann2012}. The observed enhanced $j_{r}$-regions are concentrated photospheric footprints of these currents. Flare energy release and acceleration of electrons can happen in coronal current sheets that are not related directly to the longitudinal currents \citep[e.g.][]{Priest2002,Somov2013}. Consequently, flare loop footpoints, where accelerated electrons precipitate, do not necessarily coincide with the maxima of enhanced $j_{r}$-regions.

There are quantitative differences between the results obtained on the basis of the analysis of images synthesized by CLEAN and PIXON algorithms. In particular, it is known that sizes (therefore, area) of HXR sources synthesized in a standard way using CLEAN implemented in the \textit{SolarSoftWare} (SSW), exceed values ​​obtained using other algorithms \citep{Schmahl2007,Dennis2009}. This may explain the average higher area of ​​the HXR sources constructed by CLEAN in our work (see Figure~\ref{4_2_histos_fig}). Since the areas of the HXR sources reconstructed by CLEAN and PIXON are different, other quantitative parameters of the sources, calculated for the 90\% ``core'' clusters of pixels within these areas, are also different. This applies both to the PVEC parameters (such as maximal and average absolute values of $j_{r}$, total $J_{r}$, etc.), and to the HXR fluxes. This can explain the quantitative differences in Figures~\ref{max_abs_jr_ihxrc_fig}--\ref{ratio_tot_jr_ihxrc_fig}, \ref{rh_rb_rj_fig}. Moreover, it is known that different algorithms in some cases can give different number of sources. This is typical for cases with the low signal-to-noise ratio, and also when sources have a complex shape or when several sources have very different brightness at the same time. PIXON suffers from over-resolution, if not well tuned, and, in some cases, one source can be treated as a few smaller sources \citep{Krucker2011,Felix2017}. This may explain the difference in the number of HXR sources obtained with CLEAN and PIXON for some flares indicated in Table~\ref{tab:flareinfo}. However, we emphasize that the aforementioned differences are not fundamental and do not affect strongly the main conclusion of the work on the absence of correlations between the PVEC and HXR characteristics.

One may suggest that the noise level of the constructed $j_{r}$-maps (or the threshold level $j^{thr}_{r}$ selected) is too high, and because of this some of $j_{r}$-regions were unresolved or PVECs in them were underestimated. This could be also due to the limited angular resolution ($\approx 1^{\prime\prime}$) of the HMI/SDO vector magnetograms used \citep{Barnes2018}. Another factor which could influence the results is possible variations of PVECs on time scales of the flare impulsive phase of less than 10 min. Such variations have indeed been found in several flares \citep{Tan2006,Janvier2014,Musset2015,Janvier2016,Sharykin2019}. Since we used only vector magnetograms obtained immediately before and after the impulsive phase of the flares, we do not know how PVECs changed during the impulsive phase when the HXR sources were observed. For this reason, we cannot evaluate the effect of possible variations of PVECs on the results obtained. We must also not forget that the HXR sources are mainly located in the chromosphere, while the PVECs are measured on the photosphere. The vertical (or longitudinal) current profile with height is, in general, unknown. A significant difference in currents in the photosphere and in the chromosphere is possible \citep[see ][and references therein]{Fleishman2018}. Inaccuracies could also be caused by imperfectly synthesized HXR images. All these possibilities cannot be completely ruled out at this stage. We need to wait for the next generation of instruments such as the Daniel K. Inouye Solar Telescope (DKIST), with higher spectral and angular resolution, which will be capable to measure magnetic field in different layers of the solar atmosphere, including the chromosphere. Together with new HXR telescopes, such as STIX on-board the Solar Orbiter space mission \citep{Krucker2016} or Hard X-ray Imager (HXI) preparing for the ASO-S mission \citep{Gan2019,Zhang2019a}, it will be possible to achieve further progress in understanding the role of electric currents in solar flares.


\section{CONCLUSIONS}
\label{S-Conclusions}

We performed the first statistical study of relationships between flare HXR (50--100 keV) sources and PVECs on a sample of 48 solar flares occurred in 31 different active regions in 2010--2015. Flares were chosen only on the basis of their proximity to the center of the solar disk and sufficiently high HXR fluxes above 50 keV for the synthesis of high-quality images. More than 175 HXR sources were found in 81 time intervals corresponding to the main flare HXR peaks. There are four main types of locations of the HXR sources relative to enhanced $j_{r}$-regions. Type I (23 flares, $\approx 48$\%): one or a few HXR sources overlapped with enhanced $j_{r}$-regions, while others were outside them. Type II (5 flares, $\approx 10$\%): all HXR sources were outside enhanced $j_{r}$-regions. Type III (12 flares, $25$\%): all HXR sources overlapped with enhanced $j_{r}$-regions, but not all HXR were in $j_{r}$-maxima. Type IV (8 flares, $\approx 17$\%) all HXR sources overlapped with maxima of enhanced $j_{r}$-regions.

We found that $\approx 70$\% of all HXR sources overlapped with one or a few enhanced PVEC regions, within accuracy of $\pm 3^{\prime\prime}$. However, less than $\approx 40$\% of all HXR sources overlapped with local or global maxima of enhanced PVEC regions, respectively, and less than $\approx 10$\% of the HXR sources overlapped with the major PVEC maxima of an entire parent active region. In other words, the majority ($\approx 60$\%) of the HXR sources were outside the strongest PVECs. More than in half of the flares studied there were HXR sources outside the enhanced PVEC regions. We did not find any correlations between intensity of the HXR sources and PVECs under them. We also did not find evidences of systematic decrease, i.e. dissipation of PVECs under the HXR sources during the flare impulsive phase.  

Our results confirm the previous results by \citet{deLaBeaujardiere1993}, \citet{Leka1993}, and \citet{Li1997} that the places of precipitation of accelerated electrons tend to occur in vicinity of regions with enhanced PVECs, while in majority of cases these places are outside the strongest PVECs. This fact, together with the absence of correlation between intensity of the HXR sources and PVECs under them, and together with the absence of systematic decrease of PVECs under the HXR sources, does not support the current-interruption models. This, however, does not mean that these models can be completely excluded from consideration. In particular, we cannot rule out possibility of the longitudinal currents contribution to the process of plasma heating and particle acceleration in some flares, especially in the type III and IV flares. In our opinion, such flares need to be specifically investigated in detail on the base of all available observational materials. We also expect a new generation of solar instruments to further explore the role of electric currents in the processes of flare energy release.      

\acknowledgements

We are grateful to the teams of RHESSI and HMI/SDO for the available data used in this study. We thank the anonymous reviewer for a number of useful comments. This work is supported by the Russian Science Foundation (grant No. 17-72-20134). I.V.Z. also acknowledges the Chinese Academy of Sciences (CAS) President’s International Fellowship Initiative (grant No. 2018VMB0007). W.Q.G. acknowledges funding by NNSFC (grant Nos. 114278003, U1731241 and 11921003) and by CAS (grant No. XDA15052200).


\bibliographystyle{apj}
\bibliography{bibl}

\clearpage


\startlongtable
\begin{deluxetable*}{ccccccccc}
\tablecaption{Information on the solar flares studied, the flare HXR sources and their relationship with the enhanced PVEC regions. \label{tab:flareinfo}}
\tablecolumns{9}
\tablenum{1}
\tablewidth{700pt}
\tabletypesize{\scriptsize}
\tablehead{
\colhead{N} &
\colhead{Flare} &
\colhead{GOES} &
\colhead{NOAA} & 
\colhead{RHESSI} &
\colhead{RHESSI} &
\colhead{CLEAN} &
\colhead{PIXON} &
\colhead{Type}\\
\colhead{ } &
\colhead{SOLyyyy-mm-ddThh:mm} &
\colhead{class} &
\colhead{AR} & 
\colhead{t$_{i}$, UT} &
\colhead{t$_{i+1}$, UT} &
\colhead{N HXRs} &
\colhead{N HXRs} &
\colhead{HXRs-j$_{r0}$}
}
\startdata
1 & SOL2010-10-16T19:07 & M2.9 & 11112 & 19:10:40 & 19:11:04 & $2$ & $2$ & I \\ \hline
2 & SOL2011-02-13T17:28 & M6.6 & 11158 & 17:33:44 & 17:34:04 & $2_{p}$ & $2_{p}$ & III \\
 & & & & 17:34:08 & 17:34:32 & $3$ & $1$ & \\ \hline
3 & SOL2011-02-15T01:44 & X2.2 & 11115 & 01:48:40 & 01:49:12 & $2$ & $1$ & I \\
 & & & & 01:52:56 & 01:53:16 & $4$ & $2$ &\\
 & & & & 01:53:32 & 01:53:52 & $4$ & $4$ &\\
 & & & & 01:55:04 & 01:55:32 & $2$ & $4$ &\\ \hline 
4 & SOL2011-03-09T23:13 & X1.5 & 11166 & 23:20:16 & 23:20:52 & $3$ & $2$ & I \\
 & & & & 23:21:00 & 23:21:28 & $4$ & $4$ & \\ \hline
5 & SOL2011-07-30T02:04 & M9.3 & 11261 & 02:07:32 & 02:07:56 & $2_{p}$ & $2$ & I \\
 & & & & 02:08:04 & 02:08:24 & $1$ & $1$ & \\
 & & & & 02:08:36 & 02:09:04 & $2$ & $2$ & \\ \hline
6 & SOL2011-08-03T04:29 & M1.7 & 11263 & 04:31:08 & 04:31:36 & $2_{p}$ & $2_{p}$ & III \\ \hline
7 & SOL2011-09-06T22:12 & X2.1 & 11283 & 22:18:20 & 22:18:20 & $2$ & $2$ & Ia \\
 & & & & 22:18:44 & 22:19:04 & $2_{p}$ & $3$ & \\
 & & & & 22:21:40 & 22:22:16 & $2_{p}$ & $2$ & \\
 & & & & 22:22:30 & 22:22:50 & $2$ & $2$ & \\ \hline
8 & SOL2011-09-26T05:06 & X2.1 & 11302 & 05:06:16 & 05:06:40 & $2_{p}$ & $2_{p}$ & IV \\ \hline
9 & SOL2011-12-25T20:23 & C7.7 & 11387 & 20:26:52 & 20:27:16 & $2_{p}$ & $2$ & I \\ \hline
10 & SOL2011-12-27T04:11 & C8.9 & 11386 & 04:16:04 & 04:16:36 & $2$ & $4$ & Ia \\
 & & & & 04:16:44 & 04:17:12 & $2$ & $1$ & \\ \hline
11 & SOL2012-03-09T03:22 & M6.3 & 11429 & 03:40:40 & 03:41:40 & $1$ & $1$ & II \\ \hline
12 & SOL2012-05-10T04:11 & M5.7 & 11476 & 04:15:18 & 04:15:38 & $2$ & $1$ & I \\
 & & & & 04:16:20 & 04:16:48 & $2$ & $4$ & \\ \hline
13 & SOL2012-05-10T20:20 & M1.7 & 11476 & 20:25:40 & 20:26:24 & $2_{p}$ & $2_{p}$ & III \\ \hline
14 & SOL2012-06-03T17:48 & M3.3 & 11496 & 17:53:04 & 17:53:36 & $2$ & $2$ & II \\ 
 & & & & 17:53:44 & 17:54:04 & $2$ & $2$ & \\ \hline
15 & SOL2012-07-02T19:59 & M3.8 & 11515 & 20:01:44 & 20:02:12 & $2_{p}$ & $2_{p}$ & IV \\ \hline
16 & SOL2012-07-03T03:36 & C9.9 & 11515 & 03:39:12 & 03:40:00 & $2$ & $2$ & I \\ \hline
17 & SOL2012-07-04T09:47 & M5.3 & 11515 & 09:54:44 & 09:55:08 & $2_{p}$ & $2_{p}$ & IV \\ \hline
18 & SOL2012-07-04T16:33 & M1.8 & 11513 & 16:36:00 & 16:37:32 & $1$ & $1$ & II \\ \hline
19 & SOL2012-07-05T03:25 & M4.7 & 11515 & 03:35:32 & 03:35:56 & $1$ & $1$ & IV \\ \hline
20 & SOL2012-07-05T11:39 & M6.2 & 11515 & 11:43:44 & 11:44:44 & $2_{p}$ & $2_{p}$ & IV \\ \hline
21 & SOL2012-07-06T01:37 & M2.9 & 11515 & 01:38:32 & 01:39:08 & $2$ & $2$ & III \\ \hline
22 & SOL2012-11-13T05:42 & M2.5 & 11613 & 05:47:16 & 05:48:00 & $2$ & $1$ & III \\ \hline
23 & SOL2013-02-17T15:45 & M1.9 & 11675 & 15:47:12 & 15:47:28 & $2$ & $2$ & I \\ \hline
24 & SOL2013-05-02T04:58 & M1.1 & 11731 & 05:04:40 & 05:05:00 & $2$ & $2$ & I \\ \hline
25 & SOL2013-07-08T01:13 & C9.7 & 11785 & 01:21:56 & 01:22:16 & $2$ & $1$ & I \\ \hline
26 & SOL2013-10-28T15:07 & M4.4 & 11882 & 15:10:32 & 15:11:36 & $2$ & $2$ & IV \\ \hline
27 & SOL2013-11-06T13:39 & M3.8 & 11890 & 13:43:04 & 13:43:28 & $2_{p}$ & $3$ & IV \\
 & & & & 13:43:36 & 13:43:44 & $2_{p}$ & $2_{p}$ & \\ \hline
28 & SOL2013-11-07T03:34 & M2.3 & 11890 & 03:37:52 & 03:38:28 & $2_{p}$ & $2_{p}$ & I \\
 & & & & 03:39:16 & 03:40:00 & $3$ & $3$ & \\ \hline
29 & SOL2013-11-07T14:15 & M2.4 & 11890 & 14:27:52 & 14:28:36 & $4$ & $4$ & I \\
 & & & & 14:28:48 & 14:29:24 & $2_{p}$ & $2_{p}$ & \\
 & & & & 14:29:28 & 14:30:00 & $1$ & $2$ & \\ \hline
30 & SOL2014-01-07T10:07 & M7.2 & 11944 & 10:10:56 & 10:11:24 & $2_{p}$ & $2$ & III \\
  & & & & 10:11:28 & 10:12:00 & $2_{p}$ & $4$ & \\ \hline
31 & SOL2014-03-29T17:35 & X1.0 & 12017 & 17:45:28 & 17:45:48 & $2_{p}$ & $2_{p}$ & III \\
 & & & & 17:46:20 & 17:46:40 & $2_{p}$ & $2$ & \\
& & & & 17:46:48 & 17:47:20 & $2_{p}$ & $2$ & \\ \hline
32 & SOL2014-04-18T12:31 & M7.3 & 12036 & 12:52:00 & 12:53:32 & $3$ & $3$ & I \\
 & & & & 12:53:40 & 12:54:56 & $2$ & $2$ & \\ \hline
33 & SOL2014-06-11T05:30 & M1.8 & 4197 & 05:33:32 & 05:35:04 & $1$ & $1$ & IV \\ \hline
34 & SOL2014-09-23T23:03 & M2.3 & 4580 & 23:09:08 & 23:09:48 & $2$ & $2$ & II \\ \hline
35 & SOL2014-09-24T17:45 & C7.0 & 12172 & 17:49:08 & 17:49:48 & $2_{p}$ & $2$ & III \\ \hline
36 & SOL2014-10-22T01:16 & M8.7 & 12192 & 01:38:20 & 01:38:56 & $3$ & $3$ & I \\ 
 & & & & 01:38:56 & 01:39:16 & $2$ & $2$ & \\ 
 & & & & 01:39:28 & 01:39:44 & $2$ & $2$ & \\ \hline
37 & SOL2014-10-22T14:02 & X1.6 & 12192 & 14:05:36 & 14:06:00 & $4$ & $5$ & I \\
 & & & & 14:06:16 & 14:06:44 & $4$ & $4$ &  \\  \hline 
38 & SOL2014-10-24T07:37 & M4.0 & 12192 & 07:40:44 & 07:41:12 & $3$ & $3$ & I \\ 
 & & & & 07:41:48 & 07:42:16 & $3$ & $4$ & \\ \hline
39 & SOL2014-10-24T21:07 & X3.1 & 12192 & 21:11:56 & 21:13:12 & $4$ & $5$ & I \\ \hline
40 & SOL2014-10-26T18:07 & M4.2 & 12192 & 18:08:20 & 18:08:44 & $2_{p}$ & $2$ & III \\ 
 & & & & 18:09:04 & 18:09:32 & $1$ & $1$ & \\ \hline 
41 & SOL2014-11-09T15:24 & M2.3 & 12205 & 15:28:20 & 15:29:04 & $3$ & $3$ & III \\
 & & & & 15:29:12 & 15:30:40 & $3$ & $3$ & \\ \hline
42 & SOL2015-01-03T09:40 & M1.1 & 12253 & 09:44:56 & 09:45:36 & $2$ & $2_{p}$ & III \\
 & & & & 09:45:52 & 09:46:32 & $1$ & $2$ &  \\
 & & & & 09:46:40 & 09:47:16 & $2$ & $2$ &  \\ \hline
43 & SOL2015-03-10T03:19 & M5.1 & 12297 & 03:21:00 & 03:21:16 & $2$ & $2$ & I \\
 & & & & 03:21:24 & 03:21:52 & $2$ & $2$ & \\
 & & & & 03:22:00 & 03:23:08 & $2$ & $2$ & \\ \hline
44 & SOL2015-03-10T23:46 & M2.9 & 12297 & 23:59:52 & 00:00:20 & $2$ & $4$ & I \\
 & & & & 00:00:24 & 00:00:52 & $2$ & $4$ & \\ \hline
45 & SOL2015-03-12T04:41 & M3.2 & 12297 & 04:43:00 & 04:43:36 & $2_{p}$ & $3$ & I \\
 & & & & 04:43:48 & 04:44:20 & $2$ & $2$ & \\ \hline
46 & SOL2015-03-12T21:44 & M2.7 & 12297 & 21:47:44 & 21:48:52 & $2$ & $2$ & I \\ \hline
47 & SOL2015-05-12T11:45 & C3.0 & 12345 & 11:48:20 & 11:49:04 & $2$ & $1$ & II \\ \hline
48 & SOL2015-08-22T21:19 & M3.5 & 12403 & 21:21:32 & 21:21:56 & $2_{p}$ & $2_{p}$ & III \\ \hline
\enddata
\end{deluxetable*}

\clearpage

\appendix

\section{Additional figures for all 48 flare regions studied (online materials)}

\begin{figure}[!t]
\centering
\centerline{\includegraphics[width=0.82\linewidth]{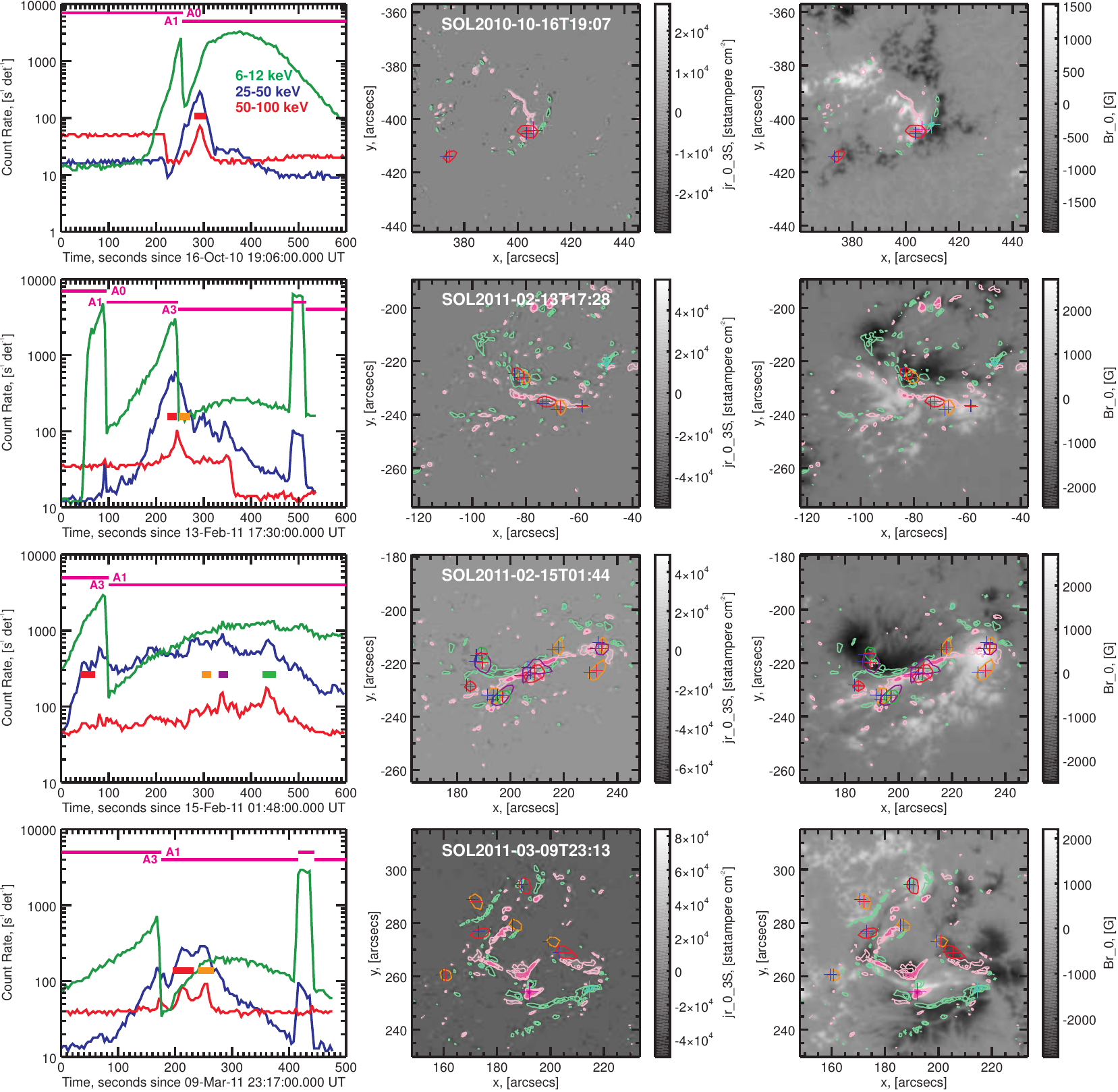}}
\caption{Lightcurves of flare HXR emission, pre-flare maps of photospheric vertical electric currents (PVECs) and magnetic field component, $B_{r0}$, with the overlying HXR sources of the solar flares No 1--4 (see Table~\ref{tab:flareinfo}). (\textit{left}) The 4-second RHESSI count rates in three energy channels 6--12 (green), 25--50 (blue), and 50--100 (red) keV. The pink horizontal lines above show the state of the RHESSI attenuators (A0, A1, A3). The thick horizontal segments of different colors (red, orange, purple, green) above the 50--100 keV count rates indicate time intervals for which images of the HXR sources (shown on the middle and right panels) were constructed. (\textit{middle}) The pre-flare maps of PVEC density, $j_{r0}$, above three standard deviations of the background, with the contour levels of $\pm 1, 2, \ldots, 8 \times j^{thr}_{r}$ (pink -- positive, cyan -- negative). The positions of the entire active region positive and negative $j_{r0}$ maxima are shown by the pink and cyan crosses, respectively. The 50--100 keV HXR sources, reconstructed with the CLEAN algorithm, at a level of 90\% of their maximum intensity, are shown by the contours of different colors corresponding to the time intervals of their appearance (shown on the left panel). The blue and red crosses show positions of centers of maximum brightness and `centers-of-mass' of brightness of the HXR sources, respectively. The sizes of the crosses indicate the estimated errors, $\pm \sigma_{\text{HXR}}$, in determining the HXR source positions. (\textit{right}) Similar to the middle panel, except that the background images on it represent the pre-flare $B_{r0}$-maps.}
\label{fig:append1a}
\end{figure}

\begin{figure}[!t]
\centering
\centerline{\includegraphics[width=0.85\linewidth]{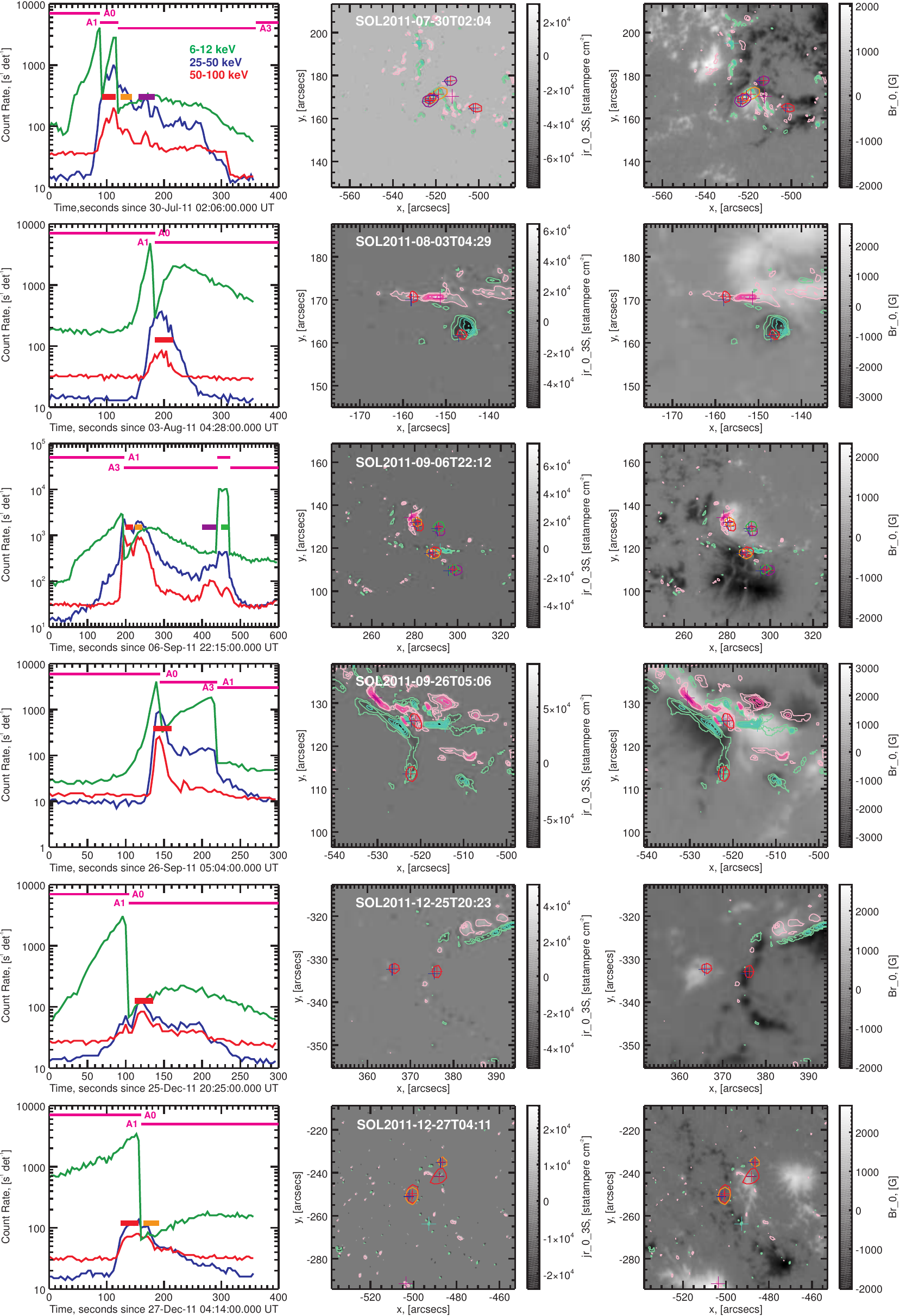}}
\caption{Same as in Figure~\ref{fig:append1a}, but for the flares No 5--10 (see Table~\ref{tab:flareinfo}). }
\label{fig:append1b}
\end{figure}

\begin{figure}[!t]
\centering
\centerline{\includegraphics[width=0.85\linewidth]{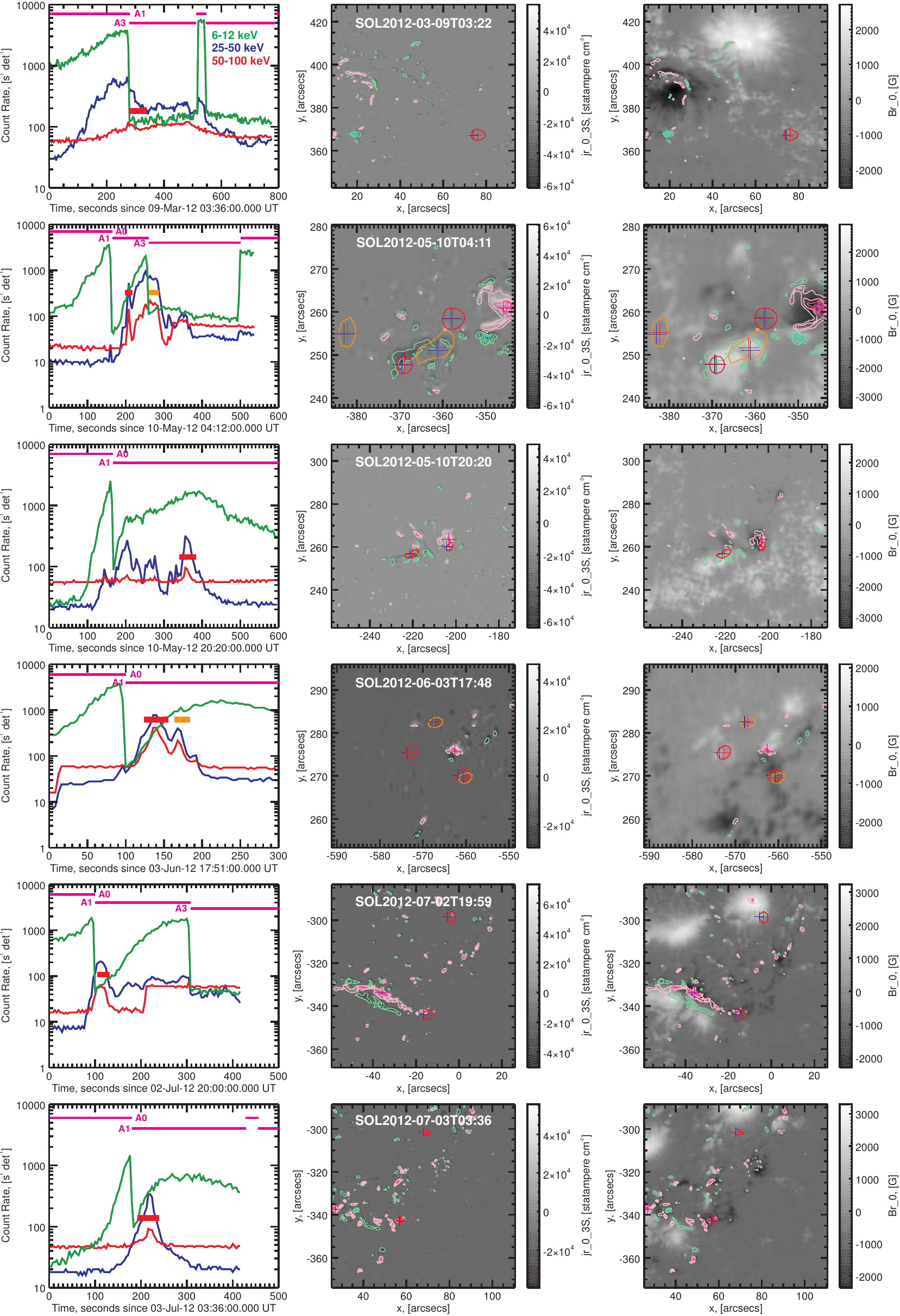}}
\caption{Same as in Figure~\ref{fig:append1a}, but for the flares No 11--16 (see Table~\ref{tab:flareinfo}). }
\label{fig:append1c}
\end{figure}

\begin{figure}[!t]
\centering
\centerline{\includegraphics[width=0.85\linewidth]{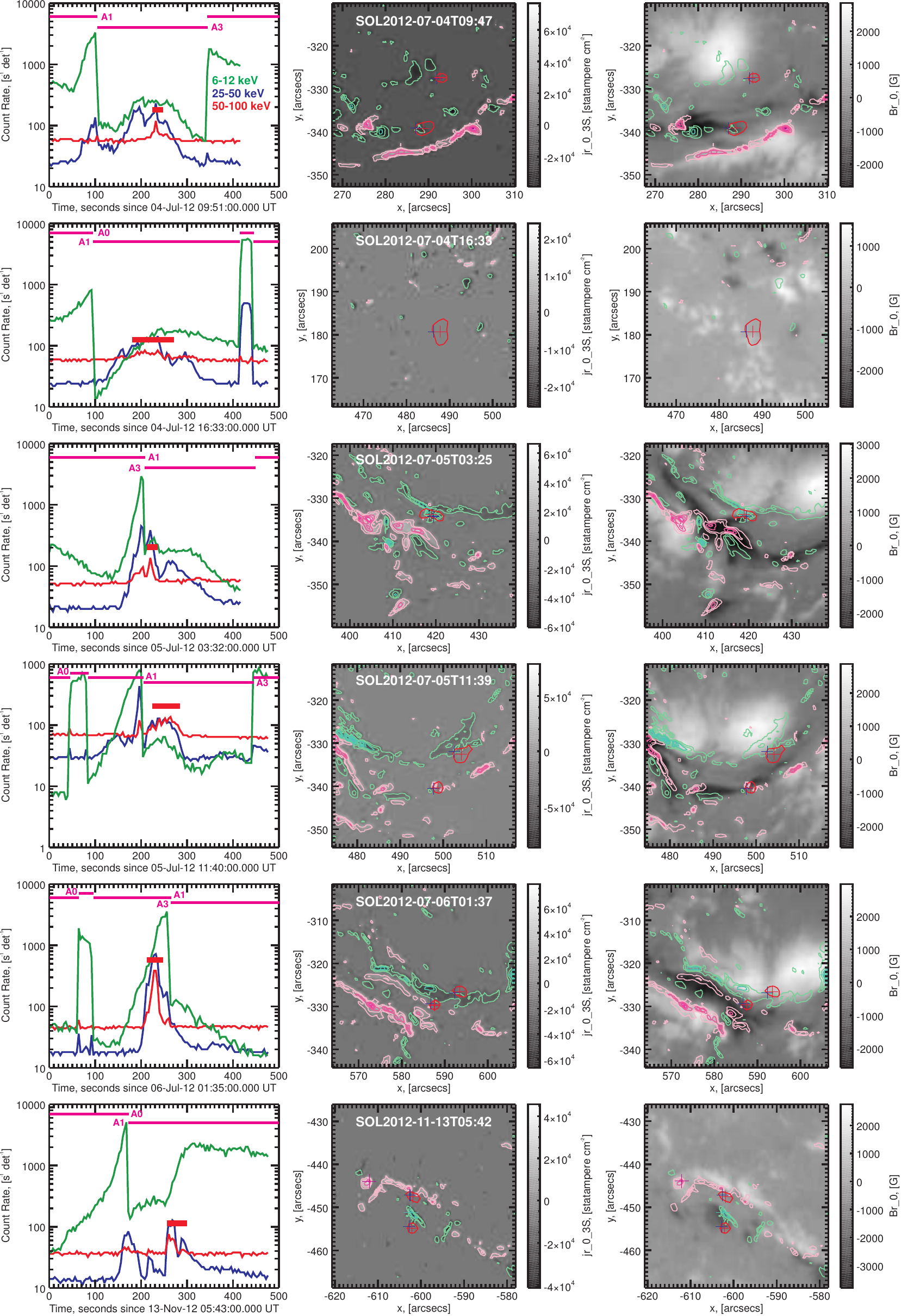}}
\caption{Same as in Figure~\ref{fig:append1a}, but for the flares No 17--22 (see Table~\ref{tab:flareinfo}). }
\label{fig:append1d}
\end{figure}

\begin{figure}[!t]
\centering
\centerline{\includegraphics[width=0.85\linewidth]{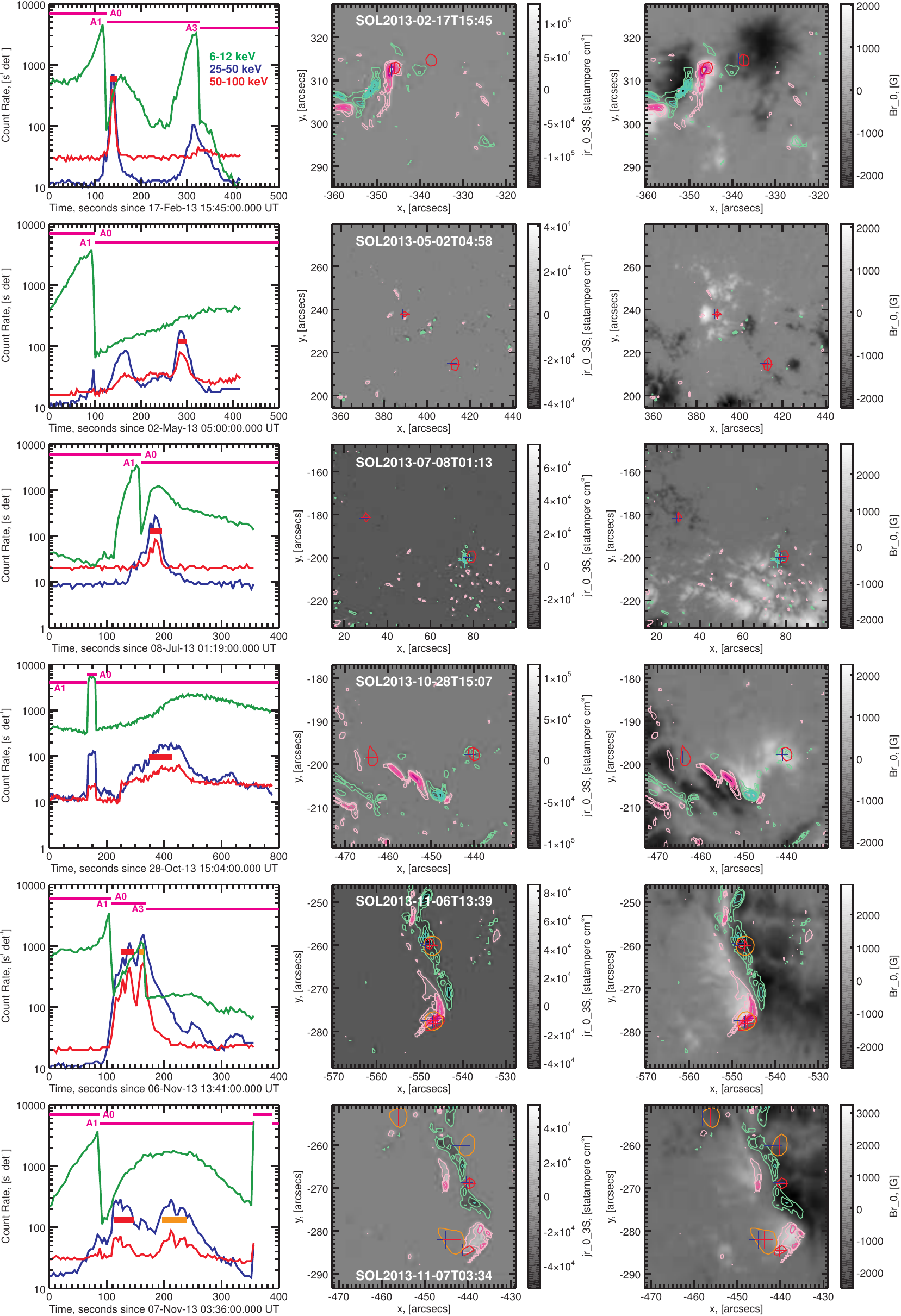}}
\caption{Same as in Figure~\ref{fig:append1a}, but for the flares No 23--28 (see Table~\ref{tab:flareinfo}). }
\label{fig:append1e}
\end{figure}

\begin{figure}[!t]
\centering
\centerline{\includegraphics[width=0.85\linewidth]{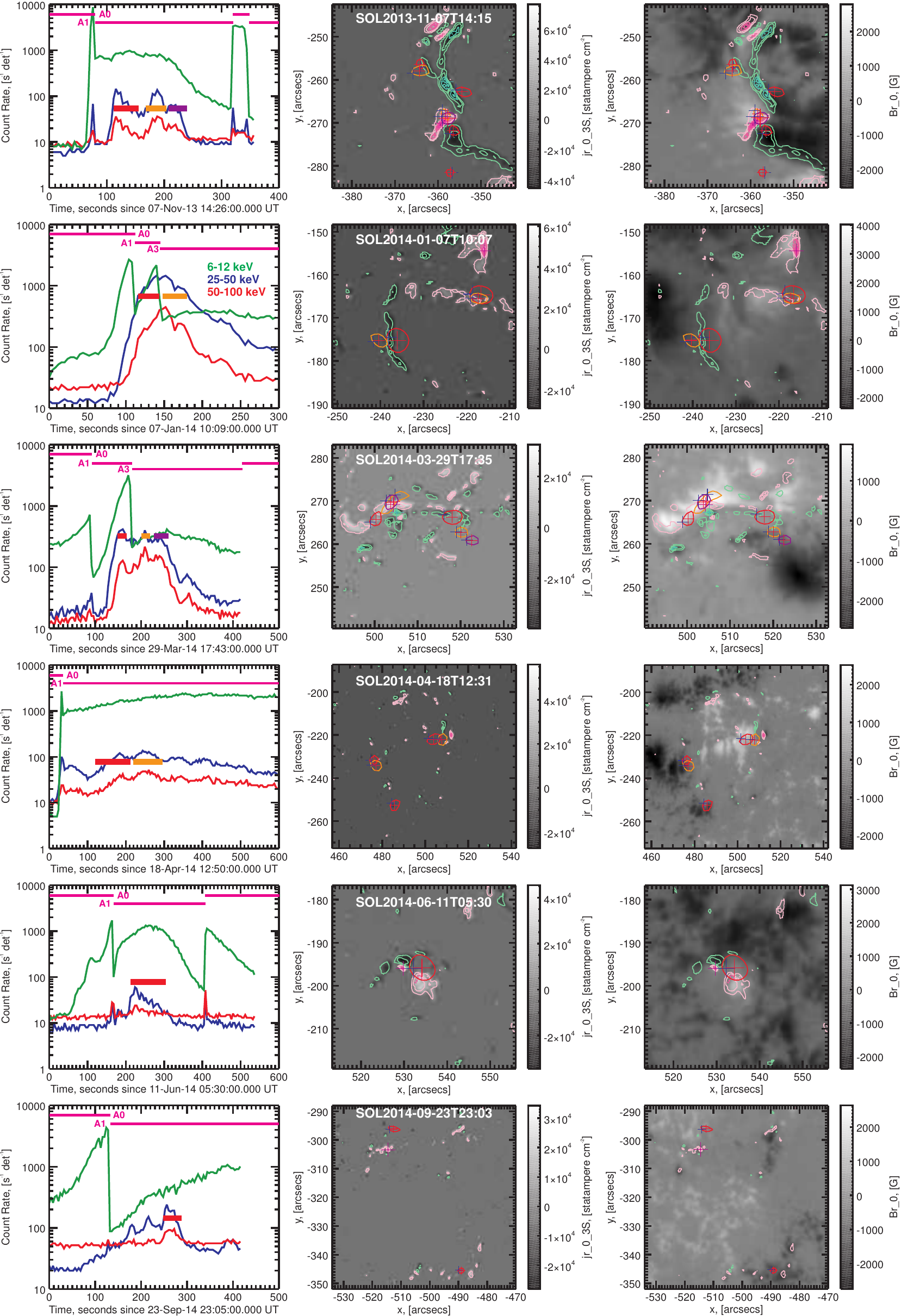}}
\caption{Same as in Figure~\ref{fig:append1a}, but for the flares No 29--34 (see Table~\ref{tab:flareinfo}). }
\label{fig:append1f}
\end{figure}

\begin{figure}[!t]
\centering
\centerline{\includegraphics[width=0.85\linewidth]{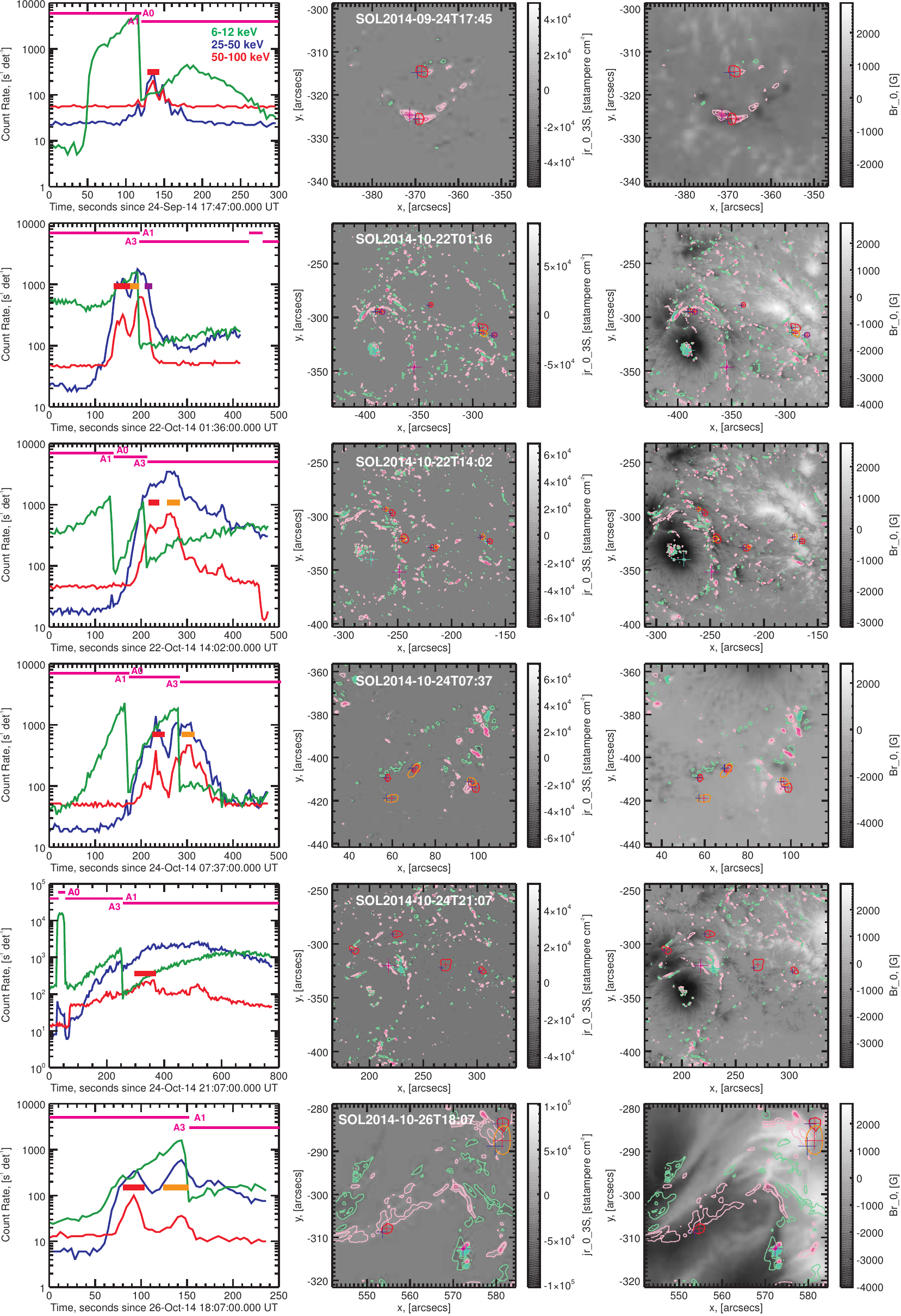}}
\caption{Same as in Figure~\ref{fig:append1a}, but for the flares No 35--40 (see Table~\ref{tab:flareinfo}). }
\label{fig:append1g}
\end{figure}

\begin{figure}[!t]
\centering
\centerline{\includegraphics[width=0.85\linewidth]{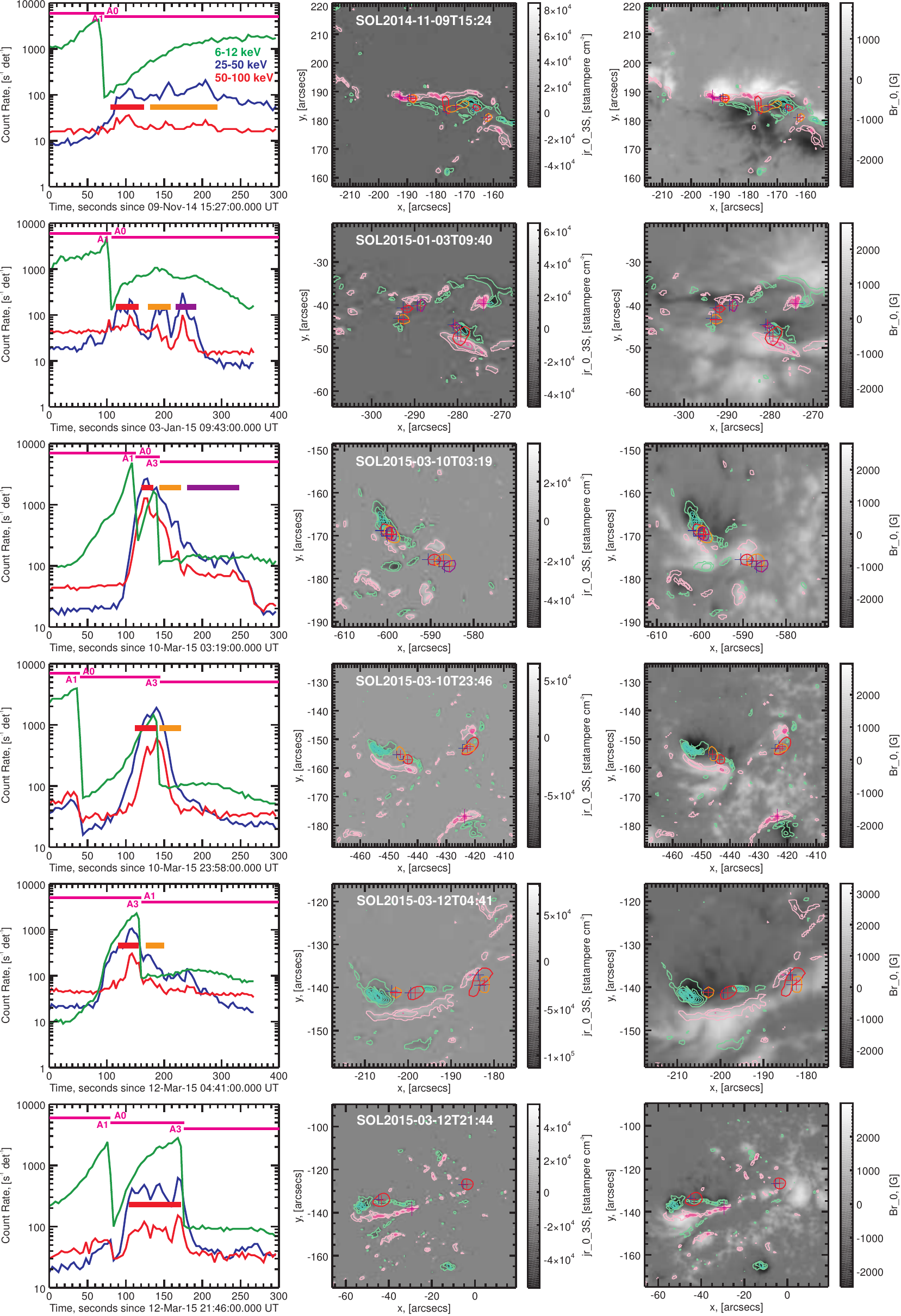}}
\caption{Same as in Figure~\ref{fig:append1a}, but for the flares No 41--46 (see Table~\ref{tab:flareinfo}). }
\label{fig:append1h}
\end{figure}

\begin{figure}[!t]
\centering
\centerline{\includegraphics[width=0.85\linewidth]{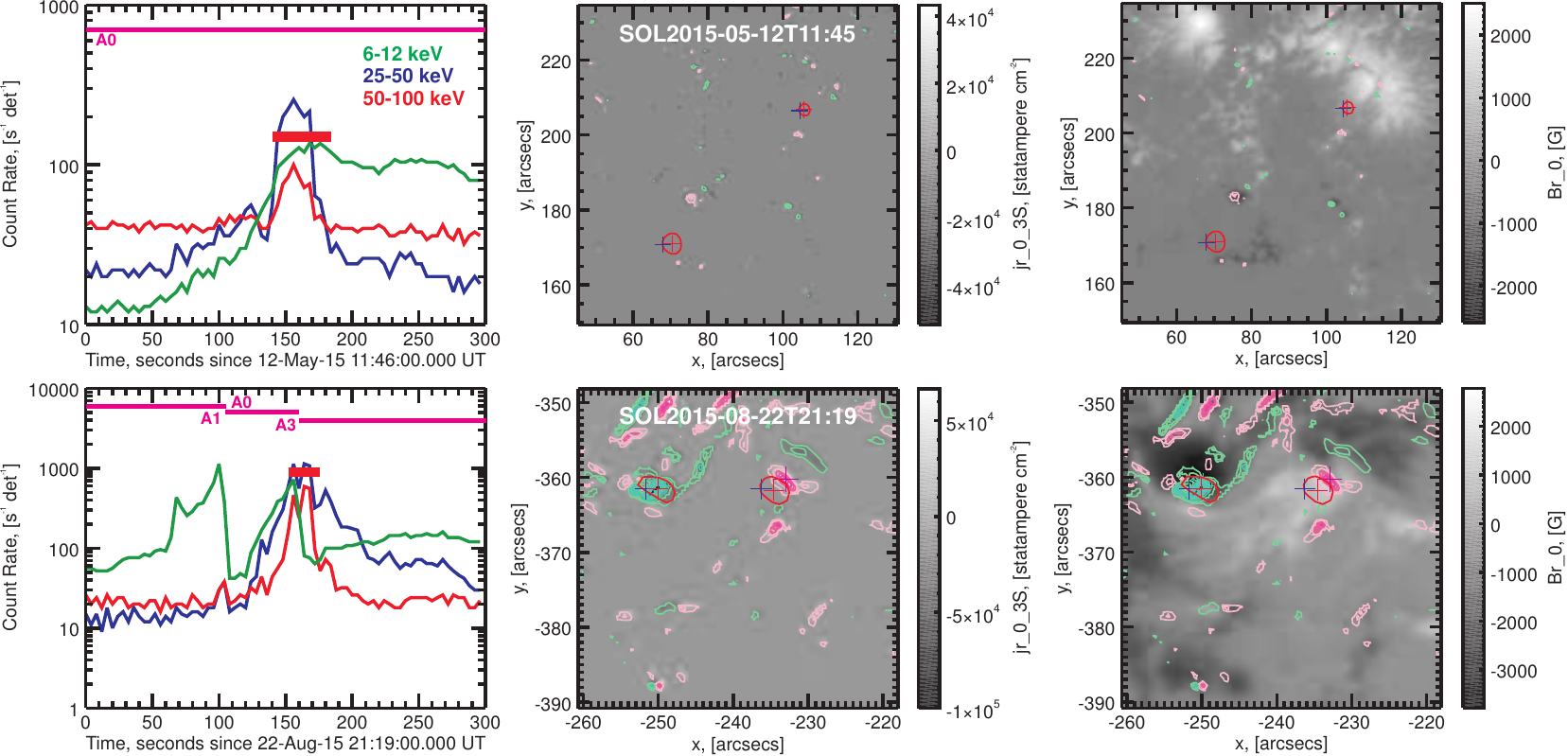}}
\caption{Same as in Figure~\ref{fig:append1a}, but for the flares No 47--48 (see Table~\ref{tab:flareinfo}). }
\label{fig:append1i}
\end{figure}

\clearpage
\end{document}